\documentclass[preprint,12pt]{elsarticle}
\usepackage[utf8]{inputenc} 
\usepackage[T1]{fontenc}
\usepackage{hyperref}
\hypersetup{
	colorlinks=true,
	linkcolor=blue,
	filecolor=magenta,
	urlcolor=blue,
}
\usepackage{mathtools}
\usepackage{amssymb}
\usepackage{amsmath}
\usepackage{graphicx}
\usepackage{latexsym}
   
\begin{document}

\journal{Journal of Geometry and Physics}

\begin{frontmatter}

\title{Eisenhart lift  of Koopman-von Neumann  mechanics}

 \author[a]{Abhijit Sen}
 \ead{abhijit913@gmail.com}
 \address[a]{Novosibirsk State University, Russia}

\author[b]{Bikram Keshari Parida}
\ead{parida.bikram90.bkp@gmail.com}  
\address[b]{Department of Physics, Pondicherry University, India}
 
\author[c]{Shailesh Dhasmana}
\ead{dx.shailesh@gmail.com}
\address[c]{University of Mons, Belgium}
 
\author[a,d]{Zurab K. Silagadze}
\ead{Z.K.Silagadze@inp.nsk.su}
\address[d]{Budker Institute of Nuclear Physics, Russia}

 \begin{abstract}
    The Eisenhart lift establishes a fascinating connection between non-relativis\-tic and relativistic physics, providing a space-time geometric understanding of non-relativistic Newtonian mechanics. What is still little known, however, is the fact that there is a Hilbert space representation of classical mechanics (also called Koopman-von Neumann mechanics) that attempts to give classical mechanics the same mathematical structure that quantum mechanics has. In this article, we geometrize the Koopman-von Newmann (KvN) mechanics using the Eisenhart toolkit. We then use a geometric view of KvN mechanics to find transformations that relate the harmonic oscillator, linear potential, and free particle in the context of KvN mechanics.
 \end{abstract}
 
 \end{frontmatter}
 
\section{Introduction}
Attempts to understand physical interactions in terms of geometry began shortly after Einstein's theory of general relativity in 1915 (although, in fact, attempts to understand mechanics geometrically have a much richer history, see, for example, \cite{Lutzen1995}). In 1928, Eisenhart proposed a variant of the geometrization of the Newtonian equations of motion of classical mechanics with $d$ degrees of freedom in terms of geodesics of a $(d+2)$-dimensional Lorentzian manifold with a Brinkmann-type metric. This formalism is called the Eisenhart lift, where the word lift indicates the necessity to introduce extra dimensions to achieve proper geometrization. However, surprisingly, Eisenhart's original publication \cite{Eisenhart} did not attract much attention and had fallen into oblivion until the idea was rediscovered in the  modern coordinate-independent approach (the so-called Bargmann structure approach) by Duval in 1985 \cite{Duval1985,Duval:1990hj,Minguzzi:2006gq}. This time these new advances have attracted attention and led to further progress in this direction, see \cite{Bekaert,Cariglia_2018,Cariglia:2016oft,Fordy:2019vzp,Zhao:2021tsz,Kan:2021yoh,Dhasmana2021} and references therein for some modern applications, including an important relation between the relativistic field equation and the Schrodinger equation \cite{Duval1985,Cariglia:2016oft}. 

Newtonian mechanics provides us with the mathematical toolkit to understand the classical world. However, the advent of quantum mechanics required a unique mathematical formalism associated with it, different from the mathematical structure used in classical physics. The change in the mathematical structure during the transition from the microscopic world (the quantum world) to the macroscopic world (the classical world) is still a mystery. Koopman-von-Newmann (KvN) mechanics arose as a result of attempts to formulate both classical and quantum mathematical formalism on the same basis of Hilbert spaces. The search for such a unified mathematical basis for both quantum and classical mechanics is still the subject of ongoing research, and two different points of view prevail in the scientific community. While some authors prefer to have a symplectic view of quantum mechanics \cite{1,2,3,4,5,6,7}, others tend to move in the direction indicated by Koopman and von Newmann \cite{9,8} of establishing a Hilbert space representation of classical mechanics \cite{Mauro,Mauro:2001rm,Gozzi:2003sh,Gozzi:2001he,Carta:2005fq,Abrikosov:2004cf}.

The Eisenhart lift of Newtonian mechanics is well known and studied. However, in this article we will develop the Eisenhart lift of the Koopman-von Neumann mechanics and point out some applications of the introduced formalism.

\section{Emergence of classicality from quantum world}
\subsection{Emergence of classical mechanics I}
Initially, classical mechanics, in the non-relativistic realm, was usually understood in the Newtonian sense. However, over time, classical mechanics has evolved from Newtonian formalism to mathematically more complex Lagrangian and Hamiltonian formulations that have given us a deep understanding of the symmetry, mathematical structures, etc. of classical dynamical systems. The Hamiltonian formalism, in particular, played a significant role in the construction of quantum mechanics and the corresponding mathematical apparatus in a Hilbert space. Both classical and quantum theories find their application in explaining the macroscopic and microscopic worlds, respectively. However, the emergence of the classical macroscopic world from the underlying microscopic quantum world is not so  straightforward to understand. In particular, the emergence of classicality from quantum mechanics is still a subject of debate \cite{Giulini:1996nw}. We would not be completely satisfied with either a naively attractive way of obtaining classicality from quantum theory using the $\hbar \rightarrow 0$ limit \cite{Padmanabhan_2015}, nor reasoning based on the Ehrenfest theorem \cite{Shankar_1994}, nor any approximate methods or arguments proposed by decoherence experts \cite{Schlosshauer:2019ewh,Ball2008}. This is due to the fact that all of them are fraught with some conceptual problems and/or problems of mathematical rigor, see, for example, \cite{Klein2012,holland_1993,Bohm_1993,Berry:1972na}.

Nevertheless, ignoring possible mathematical and conceptual subtleties, let us briefly outline the use of the $\hbar \rightarrow 0$ limit to derive classicality from a quantum theory.  It might be expected that after such a limiting procedure we would obtain Newton's equation directly from the Schrödinger equation, which, however, is not the case and we actually arrive at the Hamilton-Jacobi (HJ) equation \cite{Klein2012}. The procedure for obtaining the HJ equation includes the Madelung substitution $\psi=\sqrt{\rho}\, e^{i S/ \hbar}$ in the Schr\"{o}dinger equation and the separation of the real and imaginary parts of the resulting equation. Then, in the limit $\hbar \rightarrow 0$, we get a system whose second equation is formally identical to the Hamilton-Jacobi equation (Einstein summation convention is used):
\begin{eqnarray} &&
\frac {\partial \rho }{\partial t}+\frac {\partial }{\partial x_i}
\left( \rho v_i\right) =0, \;\;\;\; v_i=\frac{1}{m}\frac {\partial S}{\partial x_i}, \nonumber \\ &&
\frac {\partial S}{\partial t}+\frac {1}{2m}\left( \frac {\partial S}
{\partial x_i}\right) ^{2}+V\left( x,t\right)=0.
\label{eq1.1}
\end{eqnarray}
However, the system (\ref{eq1.1}) still describes the probabilistic situation of the statistical ensemble, as evidenced by the presence of the first equation, and not a Newtonian particle moving along a certain trajectory. Indeed, it can be shown \cite{Vedenyapin,Sen_2020} that if we start with the Liouville equation in phase space 
\begin{eqnarray} &&
\frac {\partial f(\vec{x},\vec{p},t)}{\partial t}=\left\{ H(\vec{x},\vec{p},t), f(\vec{x},\vec{p},t)\right\},
\nonumber \\ &&
\left\{ H, f \right\}=\frac{\partial H}{\partial x_i}\,\frac{\partial f}{\partial p_i}-\frac{\partial H}
{\partial p_i}\,\frac{\partial f}{\partial x_i},
\label{eq1.2}
\end{eqnarray}
and make the hydrodynamic substitution $f(\vec{x},\vec{p},t)=\rho( \vec{x},t)\,\delta(\vec{p}-m\vec{v}(\vec{x},t))$, one can obtain equations equivalent to the system (\ref{eq1.1}) from (\ref{eq1.2}).

To obtain a deterministic limit of (\ref{eq1.1}) and hence of Schr\"{o}dinger equation, we must assume that $\rho(\vec{x},t)$ takes the form of a delta function peaked at the Newtonian trajectory. More precisely, the procedure is as follows \cite{Klein2012}. First it can be shown that the ansatz 
\begin{equation}
\rho(\vec{x},t)=\left (\frac{1}{\pi\epsilon}\right)^{3/2}\exp{\left\{-\frac{1}{\epsilon}\left (\vec{x}-\vec{r}(t)\right )^2\right \}},
\label{eq1.3}
\end{equation}
which represents $\delta(\vec{x}-\vec{r}(t))$ in the limit $\epsilon\to 0$, is under this limit a valid solution of the first continuity equation of (\ref{eq1.1}) for arbitrary $S$.
Then Hamilton equations are obtained for averaged values $\bar x_k(t)=\int d\vec{x}\,\rho(\vec{x},t)\,x_k$ and $\bar p_k(t)=\int d\vec{x}\,\rho(\vec{x},t)\,\frac{\partial S(\vec{x},t)}{\partial x_k}$. Finally, it is shown that in the limit $\epsilon\to 0$ we have $\bar x_k(t)=r_k(t)$, $\bar p_k(t)=m\dot{r}_k(t)$, and, so it follows from the Hamilton equations that $r_k(t)$ is a Newtonian trajectory.

\subsection{Emergence of classical mechanics II}
There is a subtlety in the derivation of the classical limit, evident from the previous discussion: classicality does not arise simply as the limit $\hbar\to 0$ (or the limit $m\to\infty$, where $m$ is the mass of the particle) of the Schr\"{o}dinger equation. This subtlety was noted already by Einstein in his essay on the occasion of Born's retirement from his Chair at the University of Edinburgh in 1952 \cite{Ballentine}. Einstein considered a free particle in an energy eigenstate moving between two reflecting walls and remarked that in the classical limit $m\to\infty$ (but with the energy fixed) the quantum mechanical description does not go over into the classical description of a body with a well-defined center of mass that oscillates between the walls. Rather, in this limit, the wave function describes the behavior of an ensemble of identical bodies in the sense of classical statistical mechanics. In his answer, Born pointed out, in line with point of view \cite{Klein2012} outlined above, that one should take a sharply localized wave packet rather than an energy eigenstate before passing to the macroscopic limit \cite{Ballentine}. However,
Einstein insisted that the quantum wave function must be considered to describe an ensemble of similar systems, and cannot be considered as a complete description of an individual system, as advocated by Bohr and subsequently accepted in the canonical Copenhagen interpretation of quantum mechanics \cite{Ballentine,Einstein_1936}. From Einstein's point of view, in a proper quantum ontology, the classical description must be restored in the macroscopic limit for all solutions of the Schr\"{o}dinger equation, not just for a limited class, and he emphasized that this requires the adoption of an ensemble interpretation in which there are no difficulties regarding the classical limit \cite{Ballentine}. An interesting counterexample to Einstein's claim that the classical description must always be restored in the macroscopic limit is provided by the cryogenic version of the Weber bar (a gravitational wave detector), which must be treated quantum mechanically even though it may weigh over a ton \cite{Zurek2007}.

In fact, it is quite natural to expect that the probabilistic picture inherent in quantum mechanics in the classical limit corresponds to a similar probabilistic picture in the phase space, as indicated  by the Liouville equation. It can be easily checked that, since the Liouville equation is linear in derivatives, the square root of the phase space probability density $\psi(\vec{x},\vec{p},t)=\sqrt{f(\vec{x},\vec{p},t)}$
obeys the same Liouville equation (\ref{eq1.2}). If we assume that $\psi(\vec{x},\vec{p},t)$ is complex, then it can be viewed as a kind of wave function (in phase space) attributed to a classical particle (or system), and the Liouville equation can be rewritten in the Schr\"{o}dinger-like form
\begin{equation}
i\hbar \,\frac{\partial \psi(q, p, t)}{\partial t}=\hat{L}\, \psi(q, p, t)
\label{eq2.0}
\end{equation}
where $\hat{L}$ is the Liouville operator given by 
\begin{equation}
\hat{L}=i\hbar \{H, \,\, \}=i\hbar \left(\frac{\partial H}{\partial q} \frac{\partial }{\partial p}-\frac{\partial H}{\partial p} \frac{\partial }{\partial q}\right).  
\end{equation}
Using this fact a quantum-like Hilbert space formalism, pioneered by Koopman and von Neumann \cite{9,8}, can be developed for classical statistical mechanics. The fundamental difference between the quantum and classical cases is that \cite{Klein_2018} in the latter case, if we make the Madelung substitution for the classical wave function, we get decoupled differential equations for $\sqrt{\rho}$ and $S$, in contrast to the quantum case.

Using the fact that in the classical limit there is a mapping between the Wigner quasi-probability distribution function and the KvN wave function \cite{Bondar_2013}, it can be shown that the KvN equation naturally arises from the quantum picture in the classical limit, if we start from the representation of quantum mechanics in the phase space instead of configuration space.

The Wigner function is given by (for simplicity, one-dimensional systems are considered throughout)
\begin{equation}
    W(q, p)=\frac{1}{\sqrt{2\pi \kappa \hbar}} \int e^{i p y / \kappa \hbar} \Psi^{*}(q+\frac{y}{2},t) \Psi(q-\frac{y}{2},t) d y,
    \label{eq2.1}
\end{equation}
where the parameter $\kappa$ was introduced, $\kappa\to 0$ corresponding to the classical limit when  the coordinate and momentum operators must commute.  Note that the last condition is different from the condition $\hbar\to 0$ because $\hbar$ is also present as a multiplier of the time derivative in the Schr\"{o}dinger equation. Thus, the limit $\hbar\to 0$ means more than the criterion for the commutativity of the position and momentum operators in the classical limit, and differs from the limit $\kappa\to 0$, which ensures the commutativity of these operators \cite{Bondar_2012a}.

Let $y=\kappa \hbar \lambda_{p}$. The above equation can be re-written in terms of $\lambda_{p}$
\begin{equation}
    W(q, p)= \sqrt{\dfrac{\kappa \hbar}{2\pi }} \int e^{i  p \lambda_{p} } \Psi^{*}(q+\frac{\kappa \hbar \lambda_{p}}{2},t) \Psi(q-\frac{\kappa \hbar \lambda_{p}}{2},t) d\lambda_{p}.
    \label{eq2.2}
\end{equation}
Using the following change of variables,
\begin{equation}
u=q-\frac{\kappa \hbar \lambda_{p}}{2}, \qquad v=q+\frac{\kappa \hbar \lambda_{p}}{2},
\label{eq2.3}
\end{equation}
let us define $\rho(u,v,t)= \Psi^*(v) \Psi(u)$. Since  $\Psi(q,t)$ satisfies the time-dependent Schr\"{o}dinger equation, it can be easily checked that $\rho(u,v,t)$ satisfies the following equation
\begin{equation}
    i \kappa \hbar \,\frac{\partial}{\partial t} \rho (u,v,t)=[\hat{H}_{u}-\hat{H}_{v}]\, \rho(u,v,t),
    \label{eq2.4}
\end{equation}
where the Hamiltonians are $\hat{H}_{u}=\frac {(\hbar \kappa)^2}{2m}\,\frac{\partial^2}{\partial u^2}+V(u)$ and analogously for $\hat{H}_{v}$. 
Interestingly, equation (\ref{eq2.4}) is reminiscent of the chiral decomposition method, in which the phase space variables are chosen in such a way that the Hamiltonian of the system decomposes into the difference of the Hamiltonians of two uncoupled systems \cite{Alvarez:2007ys,Alvarez:2007fw,Zhang:2011zua}.

To represent Eq. (\ref{eq2.4}) in the representation-independent operator form, we introduce generalized Bopp pseudo-differential operators \cite{deGosson2011,Hillery1984}
\begin{equation}
\begin{split}
    \hat{u}=\hat{q}-\frac{\kappa \hbar \hat{\lambda}_{p}}{2} , \quad
    \hat{v}=\hat{q}+\frac{\kappa \hbar \hat{\lambda}_{p}}{2},\\ \hat{p}_{u}=\hat{p}+\frac{\kappa \hbar \hat{\lambda}_{x}}{2}, \quad \hat{p}_{v}=\hat{p}-\frac{\kappa \hbar \hat{\lambda}_{x}}{2},
    \end{split}
    \label{eq2.6}
\end{equation}
and assume the following commutation relations
\begin{equation}
[\hat{u},\hat{p}_u]=i\kappa \hbar, \qquad [\hat{v},\hat{p}_v]=-i\kappa \hbar.
    \label{eq2.5}
\end{equation}
In the $\kappa \rightarrow 0$ limit, Eq. (\ref{eq2.5}) and Eq. (\ref{eq2.6}) suggest that $\left[ \hat{q},\hat{p} \right]=0$. We assume that this commutativity holds even for nonzero $\kappa$ so that $\hat{q}$ and $\hat{p }$ can be considered as classical position and momentum operators.

Eq. (\ref{eq2.5}) will be ensured if, in addition to $\left[ \hat{q},\hat{p} \right]=0$, the  following set of commutation relations holds:
\begin{equation}
\begin{split}
    \left[\hat{q},\hat{\lambda}_{x} \right]=i ,\quad
    \left[\hat{p},\hat{\lambda}_{p} \right]=i , \quad
    \left[\hat{\lambda}_{p},\hat{\lambda}_{x} \right]=0.
    \end{split}
    \label{eq2.7}
\end{equation}
Further we make the identification
\begin{equation}
   \hat{ \lambda}_{x}=\frac{\hat{P}}{\hbar}, \qquad   \hat{\lambda}_{p}=-\frac{\hat{Q}}{\hbar},
     \label{eq2.8}
\end{equation}
and obtain the commutation relations with canonical pairs $(\hat q,\hat P)$ and  $(\hat Q,\hat p)$:
\begin{equation*}
\left[ \hat {Q},\hat {p}\right] =i\hbar,\;\;\left[ \hat {Q},\hat {P}\right] =0,\;\;
\left[ \hat {p},\hat {P}\right] =0,
\end{equation*}
\begin{equation}
\left[ \hat {q},\hat {Q}\right] =0,\;\; 
\left[ \hat {q},\hat {p}\right] =0,\;\;
\left[ \hat {q},\hat {P}\right]=i\hbar.
\label{com_rel}
\end{equation}

The Hamiltonian $\mathcal{H}_{qc}=\hat{H}_{u}-\hat{H}_{v}$ can be expressed in terms on operators $\hat{Q},\hat{P},\hat{q}$ and $\hat{p},$ as follows \cite{Bondar_2012a}:
\begin{equation}
\begin{split}
   \hat{\mathcal{H}}_{qc}&=\hat{H}_{u}-\hat{H}_{v}
   = \left[ \dfrac{\hat{p}_{u}^{2}}{2m}+V\left( \hat{u}\right) \right] -\left[ \dfrac{\hat{p}_{v}^{2}}{2m}+V\left( \hat{v}\right) \right] \\ &=\frac{ \kappa \hat{p} \hat{P}}{m}+ V\left( \hat{q}+\frac{\kappa \hat{Q}}{2}\right)- V\left( \hat{q}-\frac{\kappa \hat{Q}}{2}\right).
   \end{split}
\end{equation}
As a result, in the $(q,Q)$ representation (when $\hat q$ and $\hat Q$ operators are diagonal), equation \eqref{eq2.4} can be written as
 \begin{equation}
     i  \hbar \frac{\partial}{\partial t} \Psi=\left[\frac{ \hat{p} \hat{P}}{m}+\frac{1}{\kappa} V \left(q+\frac{\kappa Q}{2}\right)-\frac{1}{\kappa} V\left(q-\frac{\kappa Q}{2}\right)\right]\Psi,
     \label{eq2.11}
 \end{equation}
where $\Psi(q,Q,t)$ is proportional to $\rho(u,v,t)$ \cite{Bondar_2012a}.
 
In the limit $\kappa \rightarrow 0$, we get
\begin{equation}
     i  \hbar \frac{\partial}{\partial t} \psi_{KvN} =\left[\frac{ \hat{p} \hat{P}}{m}+\dfrac{\partial V(q)}{\partial q} Q\right]\psi_{KvN} = \hat{\mathcal{H}}_{c} \psi_{KvN},
     \label{eq2.12}
 \end{equation}
where $\hat{\mathcal{H}}_{qc}^{\kappa \rightarrow 0}=\hat{\mathcal{H}}_{c}$ and $\psi_{KvN}=\Psi(q,Q,t)$. The above equation is KvN equation in the $(q,Q)$ representation, where it is more quantum-like \cite{Sudarshan1976,Sherry1978}.

The Hamiltonian $\hat{\mathcal{H}}_{c}$ can be further re-written in terms of classical Hamiltonian $H$ and the KvN equation takes the following form  
\begin{equation}
    i  \hbar \frac{\partial}{\partial t} \psi=\hat{\mathcal{H}}_{c} \psi=\left( \frac{\partial \hat{H}}{\partial q} \hat{Q}+\frac{\partial \hat{H}}{\partial p} \hat{P} \right) \psi.
    \label{eq2.13}
\end{equation}
Note that we have removed the subscript KvN from the wave function, and therefore $\psi$ should be understood as the KvN wave function in what follows. 

In the representation $(q,p)$, according to commutation relations (\ref{com_rel}), $\hat P$ and $\hat Q$ are the following operators
\begin{equation}
\hat P=-i\hbar\,\frac{\partial}{\partial q},\;\;\;
\hat Q=i\hbar\,\frac{\partial}{\partial p},
\label{eq2.14}
\end{equation}
and we obtain the Liouville equation (\ref{eq2.0}) in the Schr\"{o}odinger-like form.  

\begin{figure}
    \centering
    \includegraphics[width=0.8\linewidth]{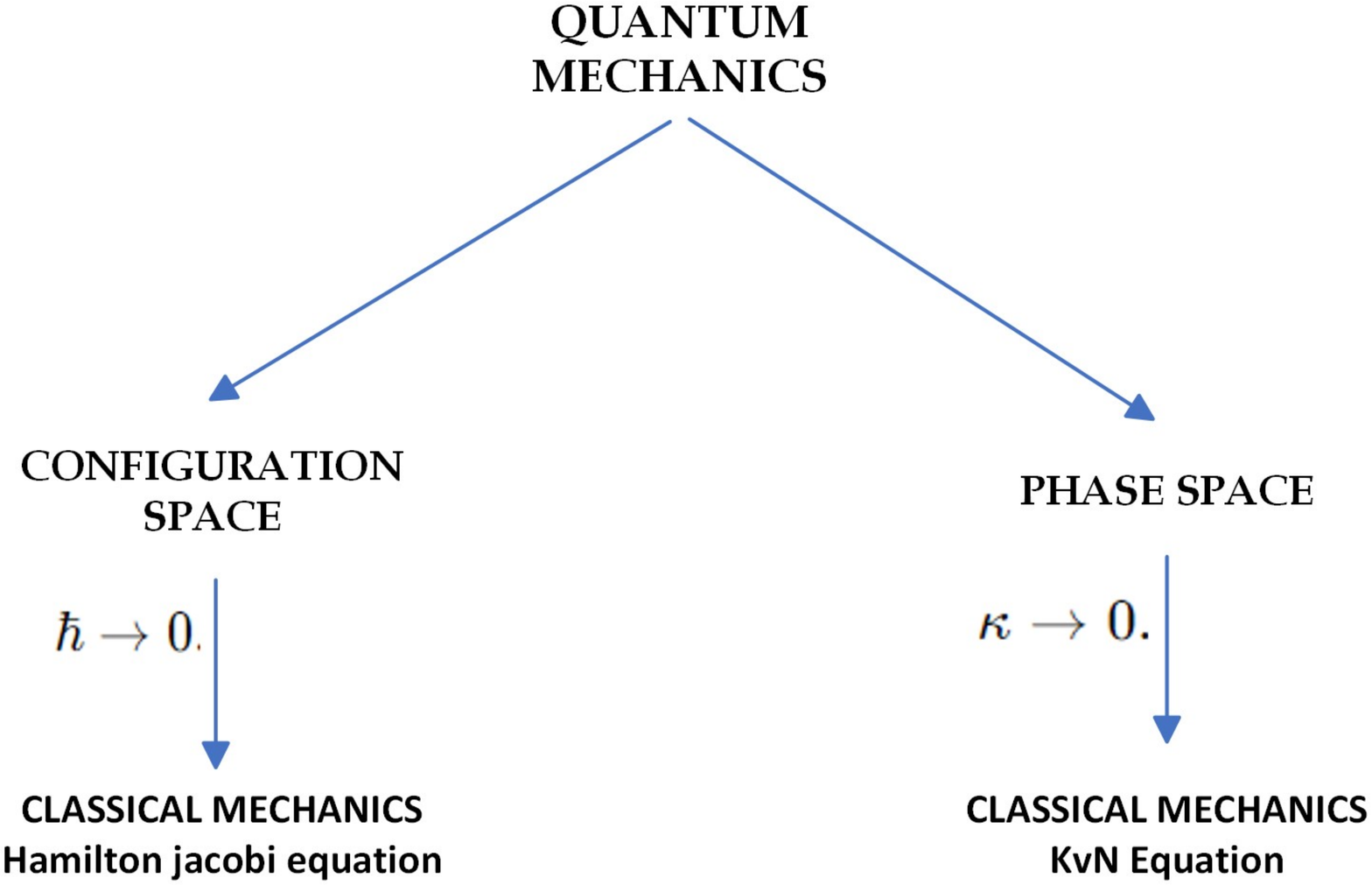}
    \caption{Emergence of KvN mechanics.}
    \label{fig:1}
\end{figure}

Therefore, the KvN formalism for the classical theory can be motivated by the quantum mechanical phase space formalism based on the Wigner quasi-probability distribution function in the $\kappa\rightarrow 0$ limit. See \cite{Sen_2020} for an alternative hydrodynamic motivation of the Madelung type, and \cite{Morgan:2019azd} for an algebraic approach to KvN mechanics. We summarize the ideas presented in this section using Fig.\ref{fig:1}. 

\subsection{Relationship between Newtonian mechanics and KvN mechanics}
A description of classical mechanics in various forms is outlined in the previous subsections. It is well known that Lagrange's equation of motion, as well as Hamilton's equation of motion lead to Newton's equation of motion. In this subsection, we will see how one can arrive at the classical Hamiltonian equation of motion starting from the KvN mechanics in the Heisenberg picture \cite{Chashchina_2020}. In the Heisenberg picture, the operators evolve as
\begin{equation}
\hat{\Lambda}(t)=e^{i \hat{L} t} \hat{\Lambda}(0) e^{-i \hat{L} t} .
\label{2.15}
\end{equation}
leading to the following equation of motion 
\begin{equation}
 \frac{d \hat{\Lambda}(t)}{d t}=i[\hat{L}, \hat{\Lambda}(t)] . 
 \label{2.16}
\end{equation}
Assuming $(q,p)$ representation, for diagonal (multiplicative) position and momentum operators we get
\begin{equation}
\frac{d q}{d t}=i[\hat{L}, \hat q]=\frac{\partial H(q, p, t)}{\partial p}, \quad \frac{d p}{d t}=i[\hat{L}, \hat p]=-\frac{\partial H(q, p, t)}{\partial q},   
\label{2.17}
\end{equation}
which are the Hamiltonian equation of motion in classical mechanics.

Another way to see that the KvN mechanics for definite initial values of $q$ and $p$ leads to Newtonian trajectories is the method of characteristics \cite{wilczek_2022}. The KvN equation (\ref{eq2.12}) is a linear partial differential equation $\psi_t+\frac{p}{m}\,\psi_q-\frac{\partial V}{\partial q}\,\psi_p=0$ , whose characteristic curves are given in parametric form by a system of ordinary differential equations
\begin{equation}
\frac{dt}{d\tau}=1,\;\;\;\frac{dq}{d\tau}=\frac{p}{m},\;\;\;\frac{dp}{d\tau}=-\frac{\partial V}{\partial q}.
\label{char_curve}
\end{equation}
The solution of (\ref{char_curve}) with initial values $q(t_0)=q_0$ and $p(t_0)=p_0$ are clearly a Newtonian trajectory $q(q_0,p_0,t)=q_N(t), \;p(q_0,p_0,t)=m\,\frac{dq_N}{dt}$. The wave function $\psi$ remains constant along characteristic curves. Therefore, if initially it is a delta-function localized around the initial values $(q_0,p_0)$, it will remain a delta-function localized around the Newtonian trajectory $(q_N(t),p_N(t))$ at every instant of time $t$.

\section{Eisenhart-lift and KvN mechanics}
\subsection{Eisenhart-lift: Brief Introduction}
Minkowski's four-dimensional (4D) representation of special relativity was an important step towards the geometrization of the theory, which further deepened our understanding of special relativity and played a crucial role in the creation of general relativity. A similar approach to classical mechanics \cite{Havas} and Newtonian theory of gravity led to the Newton-Cartan (NC) theory \cite{Cartan1923,Havas}. The geometrization of any theory introduces a mathematical structure with appropriate properties and symmetries.  For example, in the case of special relativity, the geometrization procedure introduces the Minkowski structure $(\mathcal{M},g)$, where $\mathcal{M}$ is the space-time manifold and $g$ is the Lorentzian metric associated with it. The symmetry group of this space-time is the Poincaré group $ISO(3,1)$. Similarly, NC theory introduces the Gallilean structure $(\mathcal{M},h,\tau)$, where $\mathcal{M}$ is a space-time manifold with a degenerate metric $h$ and a nowhere vanishing 1-form $\tau$. For a free particle (in 3D), the symmetry group is the Galilean group $Gal(3,1)$.

The Newton-Cartan theory, a fascinating approach to the geometrization of Newtonian gravity, is not a simple geometric theory because it has a more complex structure than Lorentzian space-time due to the presence of a degenerate metric structure: a separate metric for space, a separate metric for time, which do not merge into one actual space-time metric. In order to have a deeper understanding of the NC theory, further research led to a geometric formulation of Newtonian gravity based on gauging the Bargmann algebra (centrally extended Galilean algebra) and to a possible role of Newton-Cartan theory in non-relativistic applications of the AdS/CFT correspondence \cite{Andringa:2010it}.

Motivated by the problem of transforming the quantization of classical systems into a rigorous mathematical procedure, Alexander Kirillov, Bertram Kostant and Jean-Marie Souriau independently initiated a geometric quantization program in the 1960s \cite{Kirillov_1962,Kirillov2001,Kostant,Souriau,Souriau1}. Forming a bridge between quantum mechanics and the representation theory of Lie groups, geometric quantization provides a differential-geometric method for constructing a quantum theory corresponding to a classical system \cite{Sniatycki}. For examples of interesting interplay between geometric quantization and KvN mechanics, see, for example, \cite{Klein_2018,Abrikosov:2003ce,Bondar_2019}.

A by-product of such research was the formulation of classical mechanics in terms of Bargmann structures \cite{Duval1985,Kunzle1986}. In fact, such a geometric embedding of the classical dynamics was already established by Eisenhart \cite{Eisenhart}. However, surprisingly, at that time this publication did not attract much attention from physicists \cite{Fordy:2019vzp}.

The simplest way to explain Eisenhart lift is to use Hamiltonian approach \cite{Gibbons:2011hg,Cariglia_2015}. Given a classical dynamical system with Hamiltonian 
\begin{equation}
H =\frac{1}{2m} \sum_{i,j =1}^{n} h^{ij}(q) p_{i} p_{j} + V(q,t),
\label{class_H}
\end{equation}
the first step is to promote time $t$ to a dynamical variable (enlarge the configuration space from $\{q_i\}$ to $\{q_i,t\}$). Since
$$\int\limits_{t_i}^{t_f}L\left (q,\frac{dq}{dt},t\right )dt=\int\limits_{\sigma_i}^{\sigma_f} L\left (q,\frac{\dot{q}}{\dot{t}},t\right )\dot t\,d\sigma,$$
where $\dot{q}_i=\frac{dq_i}{d\sigma },\;\dot t=\frac{dt}{d\sigma}$, the new Lagrangian takes the form ($h_{ij}$ matrix is inverse of $h^{ij}$)
$$\tilde L(q,t,\dot{q},\dot t)=\left [\frac{m}{2} \sum_{i,j =1}^{n} h_{ij}(q) \dot{q}^{i} \dot{q}^{j}\,\dot{ t}^{-2} - V(q,t)\right ]\dot t$$
and the momentum variable canonically conjugate to time is 
$$p_t=\frac{\partial \tilde L}{\partial \dot t}=-\left [\frac{m}{2} \sum_{i,j =1}^{n} h_{ij}(q) \dot{q}^{i} \dot{q}^{j}\,\dot{t}^{-2} + V(q,t)\right ]=-H.$$
Therefore, in order to turn time into a dynamical variable, we must impose a constraint
\begin{equation}
p_t+H(q,p,t)=0,
\label{time_constraint}
\end{equation}
which ensures reparametrization invariance with respect to the auxiliary parameter $\sigma$ with the dimension of time \cite{Bekaert,Arnowitt:1962hi}.

Using $\dot{q}^i=\frac{\dot{t}}{m}\sum_{j=1}^nh^{ij}p_j$, we find $\sum_{i=1}^n\dot{q}^ip_i-\tilde L=\dot{t}H(q,p,t)$. Therefore the new Hamiltonian is
$$\tilde H=\sum_{i=1}^n\dot{q}^ip_i+\dot{t}p_t-\tilde L=
\dot{t}[p_t+H(q,p,t)].$$
Because of the reparametrization invariance, $\dot{t}$ remains arbitrary. Therefore, we can take $\dot{t}=1$ and end up with the Hamiltonian
\begin{equation}
\tilde H=p_t+H(q,p,t)=p_t+\frac{1}{2m}\sum_{i,j =1}^{n} h^{ij}p_ip_j+V(q,t).   
\label{tilde_H}
\end{equation}
Of course, this Hamiltonian vanishes due to the constraint (\ref{time_constraint}), a situation analogous to the invariance under general coordinate transformations in general relativity \cite{Arnowitt:1962hi}.

The main idea behind the Eisenhart lift is to introduce a new momentum $p_s$ conjugate to a dummy configuration space variable $s$ to make the Hamiltonian homogeneous in canonical momenta and turn it into a geodesic Hamiltonian (homogeneous quadratic function of momenta) \cite{Cariglia_2015,Gibbons:2011hg}. So we replace the Hamiltonian (\ref{tilde_H}) with the homogenized version (in the following, we use natural units with $\hbar=1$, $c=1$, and assume that $s$ has the dimension of time, so $p_s$ has the dimension of energy)
\begin{equation}
\mathcal{H} =\frac{1}{2m} \sum_{i,j=1}^{n} h^{ij}(q) p_{i} p_{j} + \frac{1}{m^2}\,p_{s}^2\, V(q,t) + \frac{1}{m}\,p_{s} p_{t}.   \label{H_Eisenhart} 
\end{equation}
Note that other conventions are also used in the literature: if $p_s$ is assumed to have the dimension of mass, then $s$ has the dimension of action per mass \cite{Duval:2014uoa}, and if $p_s$ is assumed to be dimensionless, then $s$ has the dimension of action \cite{Cariglia_2015}.

Since $s$ is a cyclic coordinate, $p_s$ is conserved and we can take $p_s=m$. In this case, (\ref{H_Eisenhart}) reduces to (\ref{tilde_H}). Thus, in fact, we have homogenized the constraint (\ref{time_constraint}) to obtain a constraint that is quadratic in momenta \cite{Bekaert}:
\begin{equation}
\mathcal{H} =\frac{1}{2m}\sum\limits_{A,B=1}^{n+2}g^{AB}p_Ap_B=0,
\label{null_H}
\end{equation}
where $p_{n+1}=p_t$, $p_{n+2}=p_s$ and the nonzero components of the inverse metric are $g^{ij}=h^{ij}$, $g^{n+1,n+2}=g^{n+2,n+1}=1$, 
$g^{n+2,n+2}=2V(q,t)/m$.

Since 
$$ \left (\begin{array}{cc} 0 & 1 \\ 1 & 2\frac{V(q,t)}{m}\end{array}\right )^{-1}=\left (\begin{array}{cc} -2\frac{V(q,t)}{m} & 1 \\ 1 & 0 \end{array}\right ),$$
the constraint (\ref{null_H}) can be interpreted as a mass-shell condition for a massless particle in space-time with a Brinkmann-type metric
\begin{equation}
dS^2 = \sum_{i,j=1}^{n} h_{ij} \,  dq^{i} \, dq^{j} + 2 ds \, dt - 2 \frac{V(q,t)}{m} \, dt^2. 
\label{eisn_metric}
\end{equation}
Newtonian trajectories in the subspace $(q_i,t)$ (space-time) are projections of zero geodesics in the Bargmann space $(q_i,t,s)$ with the Eisenhart metric (\ref{eisn_metric}) \cite{Zhang:2019gdm,Duval:2014uoa}.

There is some arbitrariness in generalizing the null constraint (\ref{time_constraint}) to (\ref{null_H}). In particular, we can multiply (\ref{null_H}) by $\Omega^{-1}(q,t)$, where the conformal factor $\Omega(q,t)$ nowhere vanishes. In this case, instead of the Eisenhart metric (\ref{eisn_metric}), we get the metric
\begin{equation}
dS^2 = \Omega(q,t)\left [\sum_{i,j=1}^{n} h_{ij} \,  dq^{i} \, dq^{j} + 2 ds \, dt - 2 \frac{V(q,t)}{m} \, dt^2\right ]. 
\label{conf_eisn_metric}
\end{equation}
The metric in this form was first introduced by Lichnerowicz \cite{Bekaert}.

\subsection{Eisenhart-lift in KvN Mechanics : Examples} 
\subsubsection{Eisenhart lift of KvN oscillator}
The study of harmonic oscillators has always been a good starting point for a deeper understanding of the physics of our world. As Sidney Coleman once said: ``The career of a young theoretical physicist consists of treating the harmonic oscillator in ever-increasing levels of abstraction" \cite{Nastase}. Therefore, we will also use the KvN harmonic oscillator to get a crude idea of the Eisenhart lift in KvN mechanics.

The equation of motion of the KvN harmonic oscillator in the $(q,Q)$-representation and assuming that the mass and spring constant are equal to unity is as follows:
\begin{equation}
    i \hbar \frac{\partial}{\partial t} \psi(q,Q,t)=( \hat{p} \hat{P} + qQ) \psi(q,Q,t).
    \label{eq3.B.1}
\end{equation}
Using the following transformation
\begin{eqnarray} &&
q =\frac{u+v}{2}, \quad\;\; Q=u-v,\nonumber \\ && 
\hat{p} =\frac{\hat{p}_{u}+\hat{p}_{v}}{2}, \;\; \hat{P}=\hat{p}_{u}-\hat{p}_{v},
\label{eq3.B.2}
\end{eqnarray}
the KvN equation of motion for harmonic oscillator becomes
\begin{equation}
     i \hbar \frac{\partial}{\partial t} \psi(u,v,t)= (\hat{H}_{u}-\hat{H}_{v})  \psi(u,v,t),
     \label{eq3.B.3}
\end{equation}
where $\hat{H}_{w}=\frac{1}{2}\left (\hat{p}_{w}^{2}+w^2\right) \text{ with } w=u,v$.

If the Hamiltonian in the equation (\ref{eq3.B.1}) is considered as the Hamiltonian of a classical system with configuration space variables $q$ and $Q$, the corresponding Eisenhart-lift metric is given by
\begin{equation}
    d S_{H O}^{2}=2 dq\,dQ+2 d t d s-2 Q q\, dt^{2},
    \label{eq3.B.4}
\end{equation}
and similar reasoning for the Hamiltonian in equation (\ref{eq3.B.3}) leads to
\begin{equation}
    d S_{H O}^{2}=d u^{2}-d v^{2}+2 d t d s-\left(u^{2}-v^{2}\right) d t^{2}.
    \label{eq3.B.5}
\end{equation}
Both Eisenhart-metric expressions (\ref{eq3.B.4}) and (\ref{eq3.B.5}) are equivalent and related by the transformation $q=\frac{u+v}{2}, \quad Q=u-v$.

Since the determinant of the matrix $$\left (\begin{array}{cc}-(u^2-v^2) & 1 \\ 1 & 0\end{array}\right )$$ is $-1 $, this matrix has one negative and one positive eigenvalue. Then from (\ref{eq3.B.5}) it is clear that this metric is an ultrahyperbolic metric, in contrast to the Eisenhart lift of ordinary classical dynamics, which gives a Lorentzian metric.

\subsubsection{Eisenhart lift of linear potential in KvN mechanics}
It follows from \eqref{eq2.12} that the equation of motion for the linear potential $V(q) = gq$ (assuming unit mass) in KvN mechanics in the ($q,Q$)-representa\-tion has the form
\begin{align}
     i h \frac{\partial}{\partial t} \psi(q,Q,t)=( \hat{p} \hat{P} +g \hat{Q})  \psi(q,Q,t).
    \label{eq1a.B.2}
\end{align}
 After using the same transformation equation \eqref{eq3.B.2}, the KvN equation of motion \eqref{eq1a.B.2} becomes
 \begin{align}
      i h \frac{\partial}{\partial t} \psi(u,v,t)= (\hat{H}_{u}-\hat{H}_{v})  \psi(u,v,t)
     \label{eq2a.B.2}
 \end{align}
Where, $\hat{H}_{w}=\frac{\hat{p}_{w}^{2}}{2}+gw \text { and } w=u, v$. As in the case of harmonic oscillator, equations \eqref{eq1a.B.2} and \eqref{eq2a.B.2} suggest two equivalent forms for the corresponding Eisenhart metric 
\begin{align}
     d S_{Linear}^{2}&=2 dq\,dQ+2 dt ds-2 gQ\, d t^{2},
    \nonumber \\
     dS_{Linear}^2 &= du^2 - dv^2 + 2 dt\,  ds - 2 g (u - v) dt^2. 
     \label{eq4a.B.2} 
\end{align}
Again, these different forms of the same Eisenhart metric are related by the transformation $q =\frac{u+v}{2}, \quad Q=u-v$, and the second form suggests that the metric is of the ultrahyperbolic signature.

\subsection{Eisenhart-lift of general KvN dynamical system}
The simplest way to geometrize the KvN mechanics is to begin from the KvN Hamiltonian that follows from (\ref{eq2.12}) and consider it as describing classical (not KvN) system:
\begin{equation}
H=\frac{pP}{m}+\frac{\partial V}{\partial q}Q.
\label{eq3.C.1}
\end{equation}
Homogenizing this Hamiltonian as described above, we get 
\begin{equation}
\mathcal{H}=\frac{pP}{m}+\frac{\partial V}{\partial q}Q\frac{p_s^2}{m^2}+\frac{1}{m}p_sp_t,
\label{eq3.C.2}
\end{equation}
which corresponds to the inverse metric
\begin{equation}
g^{qQ}=g^{Qq}=1,\;\;g^{st}=g^{ts}=1,\;\;g^{ss}=\frac{2}{m}\frac{\partial V}{\partial q}Q,
\label{eq3.C.3}
\end{equation}
all other components being zero. Inverting $g^{AB}$ to calculate the metric tensor $g_{AB}$, we get the corresponding Eisenhart metric
\begin{align}
    dS^2 = 2 dq \, dQ + 2 dt \, ds - \frac{2Q}{m}\,\frac{\partial V(q)}{\partial q} dt^2. \label{eq3.C.6}
\end{align}
Harmonic oscillator and linear potential examples considered above are just special cases of (\ref{eq3.C.6}).

However, for (\ref{eq3.C.6}) to be considered as the Eisenhart lift of the KvN dynamical system, we have to demonstrate something more: that the KvN equation (\ref{eq2.12}) can be seen as null-reduction (reduction in the $s$-direction) of the massless Klein-Gordon equation in the Bargmann space with the Eisenhart metric (\ref{eq3.C.6}).

The massless Klein-Gordon equation in general metric is given by
\begin{align}
     \square \chi&=\frac{1}{\sqrt{|g|}}  \partial_{\mu}\left(\sqrt{|g|} g^{\mu \nu} \partial_{\nu} \chi\right)=0.
     \label{eq3.C.7}
\end{align}
Using the metric in \eqref{eq3.C.6}, we get the explicit form of curved KG for the massless scalar field $\chi(t,s,q,Q)$ as follows
 \begin{align}
\dfrac{\partial^2 \chi}{\partial q \partial Q} + \frac{Q}{m}\,\frac{\partial V}{\partial q} \frac{\partial^2 \chi}{\partial s^2} + \frac{\partial^2 \chi}{\partial t \partial s}=0,
\label{eq3.C.8}
 \end{align}
which after the field redefinition 
\begin{equation}
    \chi(t,s,q,Q) = e^{ims} \psi_{KvN}(t,q,Q),
    \label{redef1}
\end{equation}
reduces to the equation of the form
\begin{align}
    i \frac{\partial \psi_{KvN}}{\partial t} = 
    \left(Q \frac{\partial V}{\partial q} - \frac{1}{m}\,\frac{\partial^{2}}{\partial q \partial Q}\right) \psi_{KvN}, 
    \label{eq3.C.9}
\end{align}
which is the KvN equation in the $(q,Q)$-representation for the  classical Hamiltonian $H = \frac{p^2}{2 m } + V(q)$,  \eqref{eq2.12} being its operator version.

Thus, the KvN equation can be considered as a null-reduction (reduction in the $s$-direction) of the Klein-Gordon equation in the Eisenhart metric background, much like the quantum mechanical case \cite{Duval1985,Horvathy:2008hd,Bekaert}. Therefore, \eqref{eq3.C.6} is indeed the Eisenhart lift of the general one-dimensional KvN system. Generalization to multidimensional KvN systems does not cause difficulties.

An interesting difference from the usual quantum mechanics is that in the KvN-Case the Eisenhart metric is of signature $(-,-,+,+)$. This implies that the free KvN-system gives rise to $SO(2,2)$ symmetric Eisenhart metric, while the free Schr\"{o}dinger-system gives rise to $SO(1,2)$ symmetric Eisenhart metric (for one-dimensional configuration space). Thus, group-theoretic analyzes like those in \cite{Minguzzi:2006wz} or \cite{Duval:2014uoa} deserve further study.

\subsection{Some applications}
An immediate application of the Eisenhart lift of KvN mechanics is to obtain relations between KvN systems with a harmonic potential, a linear potential, and a free particle. The quantum mechanical analog \cite{Dhasmana2021} hints that for this to be possible, the Eisenhart metric must be conformally flat.

In $n>3$ dimensions, conformal flatness is guaranteed by the vanishing of the Weyl tensor
\begin{align}
    W_{\mu \nu \gamma \delta} &= R_{\mu \nu \gamma \delta} - \dfrac{1}{n-2} \left(g_{\mu \gamma} R_{\nu \delta} - g_{\nu \gamma} R_{\mu \delta} - g_{\mu \delta} R_{\nu \gamma}\right. \nonumber\\
    &\left.+ g_{\nu \delta} R_{\nu \gamma} \right) + \dfrac{1}{(n-1) (n-2)} \, \left( g_{\mu \gamma} g_{\nu \delta} - g_{\nu \gamma} g_{\mu \delta}\right) \, R, 
    \label{weyl_tensor}
\end{align}
where $R_{\mu \nu \gamma \delta}$, $R_{\mu \nu }$, and $R$ are the Riemannian curvature tensor, the Ricci tensor, and the scalar curvature for the metric $g_{\mu \nu }$, respectively. For a general KvN mechanical system, using the lifted metric \eqref{eq3.C.6}  in the Weyl tensor \eqref{weyl_tensor}, we can calculate the non-zero components of the Weyl tensor as follows:
\begin{equation}
    W_{qtqt} = W_{tqtq}=-W_{qttq}=-W_{tqqt}= \, \frac{Q}{m}\,\frac{\partial^3 V(q,t)}{\partial q^3}.
\end{equation}
From the conformal flatness condition $W_{\mu \nu \gamma \delta}=0$ we obtain :
\begin{align}
    V(q,t) = \frac{1}{2} C_{1}(t) q^2 + C_{2}(t) q + C_{3}(t),
    \label{pot}
\end{align}
where $C_{1}$, $C_{2}$ and $C_{3}$ are the integration constants.

Hence we obtain a class of potentials for which the KvN Eisenhart metric \eqref{eq3.C.6} is conformally flat. The result is the same as for the Lorentzian Eisenhart lift \cite{Duval:1994qye,Dhasmana2021}: only up to quadratic-in-the-position potentials is the corresponding Eisenhart metric conformally flat.

\subsubsection{Equivalence of a harmonic oscillator to a free particle}
For the classical harmonic potential $V(q) = \frac{1}{2}q^2$, the corresponding Eisenhart metric takes the following form (we have added the Lichnerowicz conformal factor)
\begin{align}
    dS_{HO}^2 =\Omega(t) ( 2 dq \, dQ + 2 dt \, ds - 2 Q q dt^2).
\end{align}
Using the transformation $q=\frac{v+u}{2}$, $Q=u-v$, the metric is transformed into
\begin{align}
  dS_{HO}^2=\Omega(t) \left( du^{2}-dv^{2}+2dt \, ds-(u^2-v^2)dt^{2} \right).
  \label{eq5.2}
\end{align}
We know that this Eisenhart-Lichnerowicz metric is conformally equivalent to the Eisenhart metric for a free KvN particle 
\begin{align}
    dS_{Free}^2 &= d\eta^2 - d\xi^2 + 2 d\tau\,  d\zeta.
    \label{eq5.3}
\end{align}
Thus there exist a coordinate transformation $(u,v, s,t) \rightarrow (\eta, \xi, \zeta, \tau)$ between \eqref{eq5.2} and \eqref{eq5.3}. Since calculations almost exactly mirror the quantum case considered in \cite{Dhasmana2021}, we present only the final results. We choose the conformal factor $\Omega(t) = sec^{2}(t)$ and take the general coordinate transformation  as $u = \Phi(\eta,\xi,\tau), \, v = R(\eta,\xi,\tau), \, s = \zeta + \Theta(\eta,\xi,\tau), \, t = T(\eta,\xi,\tau)$. The explicit form of the transformation (with all integration constants set to zero) is 
\begin{align}
t&=\tan ^{-1}\left( \tau \right),\;\; 
u=\frac{-\eta }{\sqrt{\tau ^{2}+1}},\;\;
v=\frac{-\xi }{\sqrt{\tau ^{2}+1}},\nonumber \\
s&=\zeta +\dfrac{1}{2(\tau ^{2}+1)}\left[ \left( \eta ^{2}-\xi^{2}\right) \tau \right]. \label{eq5.4}
\end{align}
This transformation between Eisenhart metrics for KvN Harmonic oscillator and KvN free particle induce a transformation between the corresponding KvN wave functions $\psi_{HO}(u,v,t,s)$ and $\psi_{free}(\eta,\xi,\tau,\zeta)$ in the equation \eqref{eq3.C.9}:
\begin{align}
    \psi_{HO}=\sqrt{1+\tau^2} \; \exp{\left[{\dfrac{im\left[ \left(\xi^{2}-\eta^{2}\right) \tau \right]}{2(\tau ^{2}+1)}}\right]}\psi_{free}.
    \label{eq5.5}
\end{align}
One remark is pertinent here. When the metric \eqref{eq3.C.6} is multiplied by Lichnerowicz conformal factor $\Omega(t)$ and used in \eqref{eq3.C.7}, instead of \eqref{eq3.C.8} we obtain
 \begin{align}
\dfrac{\partial^2 \chi}{\partial q \partial Q} + \frac{Q}{m}\,\frac{\partial V}{\partial q} \frac{\partial^2 \chi}{\partial s^2} + \frac{\partial^2 \chi}{\partial t \partial s}+\frac{1}{2\Omega}\frac{\partial \Omega}{\partial t}\frac{\partial \chi}{\partial s}=0,
\label{eq3.C.8m}
 \end{align}
and to get the KvN equation \eqref{eq3.C.9} we need to replace \eqref{redef1} by
\begin{align}
    \chi(t,s,q,Q) = \Omega^{-1/2}e^{ims} \psi_{KvN}(t,q,Q).
    \label{redef2}
\end{align}
This explains the $\sqrt{1+\tau^2}$ prefactor before the exponent in \eqref{eq5.5}. Because of this time-dependent prefactor, the transformation \eqref{eq5.5} is not unitary, and the corresponding mapping of a harmonic oscillator into a free particle cannot be considered as an equivalence principle, in contrast to the case of a linear potential (see below). This situation is completely analogous to the quantum case \cite{Dhasmana2021}. Note, however, that there is a broader ``conformal" context in which the correspondence of a harmonic oscillator to a free particle fits \cite{Gibbons:2014zla,Zhang:2021ssp}.

The connection between quantum isotropic oscillators with a time-inde\-pen\-dent frequency and a free particle is usually attributed to Niederer \cite{Niederer:1973tz}. However, apparently, for the first time such a connection was noticed in the field of optics \cite{Yariv,Steuernagel2014} and explicitly or implicitly rediscovered by many authors \cite{Solovev:1982,Jackiw:1980mm,Guerrero:2013mdt}. This relationship has recently been discussed in the context of KvN mechanics \cite{McCaul:2022cyl}, and extended to anisotropic oscillators with time-dependent frequency \cite{Zhang:2021ssp}.

\subsubsection{Equivalence of a linear potential to a free particle}
A similar analysis can also be carried out for a linear potential. The Eisenhart metric for the KvN  linear potential $V=gq$ (assuming unit mass) has the form \eqref{eq4a.B.2}. It is converted to the free metric \eqref{eq5.3} by the following transformation (integration constants are again set to zero)
\begin{align}
    t = \tau, \;
    u = \eta - \frac{1}{2} g \tau^{2}, \;
    v = \xi - \frac{1}{2} g \tau^{2}, \;
    s = \zeta + (\eta-\xi) g \tau.
    \label{eq5.7}
\end{align}
The corresponding transformation of the KvN wave function  
\begin{align}
    \psi_{Linear}=\exp{\left[i(\xi - \eta) g \tau \right]}\psi_{free},
    \label{KvN_EP}
\end{align}
is unitary and represents Einstein's principle of equivalence in KvN mechanics.

The relation \eqref{KvN_EP} in $(q,Q)$ coordinates reads
\begin{align}
    \psi_{Linear}(q,Q,t)=e^{-iQ g \tau}\psi_{free}\left (q+\frac{1}{2}gt^2,Q,t\right ),
    \label{eq5.8}
\end{align}
which is the result obtained in \cite{Sen_2020} with a different method.

\section{Conclusions}
KvN mechanics, a Hilbert space interpretation of classical mechanics, naturally appears as a limiting situation of the Wigner function formalism \cite{Bondar_2013}. Eisenhart lift of the KvN Hamiltonian opens the way to a geometric understanding of the KvN mechanics, since the massless Klein-Gordon equation in curved space for the metric (\ref{eq3.C.6}) at null-reduction gives exactly the KvN equation. Using this geometric picture, we found the transformations of coordinates, as well as KvN wave functions, that relate KvN harmonic oscillator or the linear potential to a KvN free particle. In the case of linear potential, we recover Einstein's equivalence principle in KvN mechanics first obtained in \cite{Sen_2020}.

Eisenhart lift of KvN mechanics is very similar to the similar lift of quantum mechanics \cite{Duval1985,Horvathy:2008hd,Bekaert}. This is not surprising, since the KvN system (\ref{eq2.12}) can be interpreted as a kind of quantum system with a Hamiltonian linearly dependent on the classically ``hidden" variables $Q$ and $P$, as a result of which the classical dynamics is induced in the subspace $(q,p)$.

An interesting difference from the quantum case is that the Eisenhart metric in the KvN case has an ultrahyperbolic rather than a Lorentzian signature. This is because the kinetic energy in the KvN Hamiltonian, since it must be linear in ``hidden" momenta, is pseudo-Euclidean rather than positive definite Euclidean. Such an unusual kinetic energy in the context of classical mechanics was first considered in 1935 by Drach in search of potentials that admit a cubic integral of motion with respect to momenta \cite{drach}. References  \cite{Galajinsky:2017qsb,Cariglia:2015fva,Filyukov:2015qna} consider the Eisenhart lift of such classical systems and show that a metric with an ultrahyperbolic signature arises.

Metrics with the ultrahyperbolic signature $(-,-,+,+)$ are special cases of the so-called Kleinian space-times with the complete symmetry between space and time, which are relatively unknown to physicists \cite{Barrett:1993yn}, maybe because it is widely believed that ultrahyperbolic space-times are not deterministic in a physically meaningful sense \cite{Tegmark:1997jg}. Nevertheless, such space-times, in addition to the Eisenhart lift of the Drach-type classical mechanical and and KvN systems, arise in a number of different physical contexts, see for example \cite{Barrett:1993yn,Dehdashti:2021vmz,Alves-Junior:2020nva,Gibbons:2020nzu,Bars:2000qm,Sakharov:1984csx,Smolyaninov:2010tpq,Figueiredo:2016xfc}.

\section*{Acknowledgments}
The authors would like to thank Peter Horvathy and Anton Galajinsky for useful comments. The work of Z.K.S is supported by the Ministry of Education and Science of the Russian Federation.

\bibliographystyle{elsarticle-num-names}
\bibliography{refkvn}

\begin{thebibliography}{97}
\expandafter\ifx\csname natexlab\endcsname\relax\def\natexlab#1{#1}\fi
\providecommand{\url}[1]{\texttt{#1}}
\providecommand{\href}[2]{#2}
\providecommand{\path}[1]{#1}
\providecommand{\DOIprefix}{doi:}
\providecommand{\ArXivprefix}{arXiv:}
\providecommand{\URLprefix}{URL: }
\providecommand{\Pubmedprefix}{pmid:}
\providecommand{\doi}[1]{\href{http://dx.doi.org/#1}{\path{#1}}}
\providecommand{\Pubmed}[1]{\href{pmid:#1}{\path{#1}}}
\providecommand{\bibinfo}[2]{#2}
\ifx\xfnm\relax \def\xfnm[#1]{\unskip,\space#1}\fi
\bibitem[{L{\"u}tzen(1995)}]{Lutzen1995}
\bibinfo{author}{J.~L{\"u}tzen},
\newblock \bibinfo{title}{Interactions between mechanics and differential
  geometry in the 19th century},
\newblock \bibinfo{journal}{Arch. Hist. Exact Sci.} \bibinfo{volume}{49}
  (\bibinfo{year}{1995}) \bibinfo{pages}{1--72}. \URLprefix
  \url{https://doi.org/10.1007/BF00374699}. \DOIprefix\doi{10.1007/BF00374699}.
\bibitem[{Eisenhart(1928)}]{Eisenhart}
\bibinfo{author}{L.~P. Eisenhart},
\newblock \bibinfo{title}{Dynamical {Trajectories} and {Geodesics}},
\newblock \bibinfo{journal}{The Annals of Mathematics} \bibinfo{volume}{30}
  (\bibinfo{year}{1928}) \bibinfo{pages}{591--606}. \URLprefix
  \url{https://www.jstor.org/stable/1968307?origin=crossref}.
  \DOIprefix\doi{10.2307/1968307}.
\bibitem[{Duval et~al.(1985)Duval, Burdet, Kunzle, and Perrin}]{Duval1985}
\bibinfo{author}{C.~Duval}, \bibinfo{author}{G.~Burdet}, \bibinfo{author}{H.~P.
  Kunzle}, \bibinfo{author}{M.~Perrin},
\newblock \bibinfo{title}{Bargmann structures and newton-cartan theory},
\newblock \bibinfo{journal}{Physical Review D} \bibinfo{volume}{31}
  (\bibinfo{year}{1985}) \bibinfo{pages}{1841--1853}. \URLprefix
  \url{https://link.aps.org/doi/10.1103/PhysRevD.31.1841}.
  \DOIprefix\doi{10.1103/PhysRevD.31.1841}.
\bibitem[{Duval et~al.(1991)Duval, Gibbons, and Horvathy}]{Duval:1990hj}
\bibinfo{author}{C.~Duval}, \bibinfo{author}{G.~W. Gibbons},
  \bibinfo{author}{P.~Horvathy},
\newblock \bibinfo{title}{{Celestial mechanics, conformal structures and
  gravitational waves}},
\newblock \bibinfo{journal}{Phys. Rev. D} \bibinfo{volume}{43}
  (\bibinfo{year}{1991}) \bibinfo{pages}{3907--3922}. \URLprefix
  \url{https://doi.org/10.1103/PhysRevD.43.3907}.
  \DOIprefix\doi{10.1103/PhysRevD.43.3907}.
  \href{http://arxiv.org/abs/hep-th/0512188}{{\tt arXiv:hep-th/0512188}}.
\bibitem[{Minguzzi(2007)}]{Minguzzi:2006gq}
\bibinfo{author}{E.~Minguzzi},
\newblock \bibinfo{title}{{Eisenhart's theorem and the causal simplicity of
  Eisenhart's spacetime}},
\newblock \bibinfo{journal}{Class. Quant. Grav.} \bibinfo{volume}{24}
  (\bibinfo{year}{2007}) \bibinfo{pages}{2781--2808}. \URLprefix
  \url{https://doi.org/10.1088/0264-9381/24/11/002}.
  \DOIprefix\doi{10.1088/0264-9381/24/11/002}.
  \href{http://arxiv.org/abs/gr-qc/0612014}{{\tt arXiv:gr-qc/0612014}}.
\bibitem[{Bekaert and Morand(2013)}]{Bekaert}
\bibinfo{author}{X.~Bekaert}, \bibinfo{author}{K.~Morand},
\newblock \bibinfo{title}{Embedding nonrelativistic physics inside a
  gravitational wave},
\newblock \bibinfo{journal}{Physical Review D} \bibinfo{volume}{88}
  (\bibinfo{year}{2013}) \bibinfo{pages}{063008}. \URLprefix
  \url{https://link.aps.org/doi/10.1103/PhysRevD.88.063008}.
  \DOIprefix\doi{10.1103/PhysRevD.88.063008}.
\bibitem[{Cariglia et~al.(2018)Cariglia, Galajinsky, Gibbons, and
  Horvathy}]{Cariglia_2018}
\bibinfo{author}{M.~Cariglia}, \bibinfo{author}{A.~Galajinsky},
  \bibinfo{author}{G.~W. Gibbons}, \bibinfo{author}{P.~A. Horvathy},
\newblock \bibinfo{title}{Cosmological aspects of the eisenhart-duval lift},
\newblock \bibinfo{journal}{Eur. Phys. J. C} \bibinfo{volume}{78}
  (\bibinfo{year}{2018}). \URLprefix
  \url{https://doi.org/10.1140/epjc/s10052-018-5789-x}.
  \DOIprefix\doi{10.1140/epjc/s10052-018-5789-x}.
\bibitem[{Cariglia et~al.(2016)Cariglia, Duval, Gibbons, and
  Horvathy}]{Cariglia:2016oft}
\bibinfo{author}{M.~Cariglia}, \bibinfo{author}{C.~Duval},
  \bibinfo{author}{G.~W. Gibbons}, \bibinfo{author}{P.~A. Horvathy},
\newblock \bibinfo{title}{{Eisenhart lifts and symmetries of time-dependent
  systems}},
\newblock \bibinfo{journal}{Annals Phys.} \bibinfo{volume}{373}
  (\bibinfo{year}{2016}) \bibinfo{pages}{631--654}. \URLprefix
  \url{https://doi.org/10.1016/j.aop.2016.07.033}.
  \DOIprefix\doi{10.1016/j.aop.2016.07.033}.
  \href{http://arxiv.org/abs/1605.01932}{{\tt arXiv:1605.01932}}.
\bibitem[{Fordy and Galajinsky(2019)}]{Fordy:2019vzp}
\bibinfo{author}{A.~P. Fordy}, \bibinfo{author}{A.~Galajinsky},
\newblock \bibinfo{title}{{Eisenhart Lift of $2$-Dimensional Mechanics}},
\newblock \bibinfo{journal}{Eur. Phys. J. C} \bibinfo{volume}{79}
  (\bibinfo{year}{2019}) \bibinfo{pages}{301}. \URLprefix
  \url{https://doi.org/10.1140/epjc/s10052-019-6812-6}.
  \DOIprefix\doi{10.1140/epjc/s10052-019-6812-6}.
  \href{http://arxiv.org/abs/1901.03699}{{\tt arXiv:1901.03699}}.
\bibitem[{Zhao et~al.(2021)Zhao, Zhang, and Horvathy}]{Zhao:2021tsz}
\bibinfo{author}{Q.~Zhao}, \bibinfo{author}{P.~Zhang}, \bibinfo{author}{P.~A.
  Horvathy},
\newblock \bibinfo{title}{{Time-Dependent Conformal Transformations and the
  Propagator for Quadratic Systems}},
\newblock \bibinfo{journal}{Symmetry} \bibinfo{volume}{13}
  (\bibinfo{year}{2021}) \bibinfo{pages}{1866}. \URLprefix
  \url{https://doi.org/10.3390/sym13101866}.
  \DOIprefix\doi{10.3390/sym13101866}.
  \href{http://arxiv.org/abs/2105.07374}{{\tt arXiv:2105.07374}}.
\bibitem[{Kan et~al.(2021)Kan, Aoyama, Hasegawa, and Shiraishi}]{Kan:2021yoh}
\bibinfo{author}{N.~Kan}, \bibinfo{author}{T.~Aoyama},
  \bibinfo{author}{T.~Hasegawa}, \bibinfo{author}{K.~Shiraishi},
\newblock \bibinfo{title}{{Eisenhart-Duval lift for minisuperspace quantum
  cosmology}},
\newblock \bibinfo{journal}{Phys. Rev. D} \bibinfo{volume}{104}
  (\bibinfo{year}{2021}) \bibinfo{pages}{086001}. \URLprefix
  \url{https://doi.org/10.1103/PhysRevD.104.086001}.
  \DOIprefix\doi{10.1103/PhysRevD.104.086001}.
  \href{http://arxiv.org/abs/2105.09514}{{\tt arXiv:2105.09514}}.
\bibitem[{Dhasmana et~al.(2021)Dhasmana, Sen, and Silagadze}]{Dhasmana2021}
\bibinfo{author}{S.~Dhasmana}, \bibinfo{author}{A.~Sen}, \bibinfo{author}{Z.~K.
  Silagadze},
\newblock \bibinfo{title}{Equivalence of a harmonic oscillator to a free
  particle and eisenhart lift},
\newblock \bibinfo{journal}{Annals of Physics} \bibinfo{volume}{434}
  (\bibinfo{year}{2021}) \bibinfo{pages}{168623}. \URLprefix
  \url{https://doi.org/10.1016\%2Fj.aop.2021.168623}.
  \DOIprefix\doi{10.1016/j.aop.2021.168623}.
\bibitem[{Ashtekar and Schilling(1997)}]{1}
\bibinfo{author}{A.~Ashtekar}, \bibinfo{author}{T.~A. Schilling},
  \bibinfo{title}{Geometrical formulation of quantum mechanics},
  \bibinfo{year}{1997}. \URLprefix \url{https://arxiv.org/abs/gr-qc/9706069}.
  \DOIprefix\doi{10.48550/ARXIV.GR-QC/9706069}.
\bibitem[{Anandan and Aharonov(1990)}]{2}
\bibinfo{author}{J.~Anandan}, \bibinfo{author}{Y.~Aharonov},
\newblock \bibinfo{title}{Geometry of quantum evolution},
\newblock \bibinfo{journal}{Phys. Rev. Lett.} \bibinfo{volume}{65}
  (\bibinfo{year}{1990}) \bibinfo{pages}{1697--1700}. \URLprefix
  \url{https://link.aps.org/doi/10.1103/PhysRevLett.65.1697}.
  \DOIprefix\doi{10.1103/PhysRevLett.65.1697}.
\bibitem[{Anandan(1991)}]{3}
\bibinfo{author}{J.~Anandan},
\newblock \bibinfo{title}{A geometric approach to quantum mechanics},
\newblock \bibinfo{journal}{Foundations of Physics} \bibinfo{volume}{21}
  (\bibinfo{year}{1991}) \bibinfo{pages}{1265--1284}. \URLprefix
  \url{http://link.springer.com/10.1007/BF00732829}.
  \DOIprefix\doi{10.1007/BF00732829}.
\bibitem[{Heslot(1985)}]{4}
\bibinfo{author}{A.~Heslot},
\newblock \bibinfo{title}{Quantum mechanics as a classical theory},
\newblock \bibinfo{journal}{Phys. Rev. D} \bibinfo{volume}{31}
  (\bibinfo{year}{1985}) \bibinfo{pages}{1341--1348}. \URLprefix
  \url{https://link.aps.org/doi/10.1103/PhysRevD.31.1341}.
  \DOIprefix\doi{10.1103/PhysRevD.31.1341}.
\bibitem[{Rowe et~al.(1980)Rowe, Ryman, and Rosensteel}]{5}
\bibinfo{author}{D.~J. Rowe}, \bibinfo{author}{A.~Ryman},
  \bibinfo{author}{G.~Rosensteel},
\newblock \bibinfo{title}{Many-body quantum mechanics as a symplectic dynamical
  system},
\newblock \bibinfo{journal}{Phys. Rev. A} \bibinfo{volume}{22}
  (\bibinfo{year}{1980}) \bibinfo{pages}{2362--2373}. \URLprefix
  \url{https://link.aps.org/doi/10.1103/PhysRevA.22.2362}.
  \DOIprefix\doi{10.1103/PhysRevA.22.2362}.
\bibitem[{Marsden and Ratiu(1999)}]{6}
\bibinfo{author}{J.~E. Marsden}, \bibinfo{author}{T.~S. Ratiu},
  \bibinfo{title}{Introduction to Mechanics and Symmetry},
  \bibinfo{publisher}{Springer}, \bibinfo{address}{New York},
  \bibinfo{year}{1999}. \URLprefix
  \url{https://doi.org/10.1007\%2F978-0-387-21792-5}.
  \DOIprefix\doi{10.1007/978-0-387-21792-5}.
\bibitem[{Marmo and Volkert(2010)}]{7}
\bibinfo{author}{G.~Marmo}, \bibinfo{author}{G.~F. Volkert},
\newblock \bibinfo{title}{Geometrical description of quantum
  mechanics-transformations and dynamics},
\newblock \bibinfo{journal}{Physica Scripta} \bibinfo{volume}{82}
  (\bibinfo{year}{2010}) \bibinfo{pages}{038117}. \URLprefix
  \url{https://doi.org/10.1088\%2F0031-8949\%2F82\%2F03\%2F038117}.
  \DOIprefix\doi{10.1088/0031-8949/82/03/038117}.
\bibitem[{Koopman(1931)}]{9}
\bibinfo{author}{B.~O. Koopman},
\newblock \bibinfo{title}{Hamiltonian {Systems} and {Transformation} in
  {Hilbert} {Space}},
\newblock \bibinfo{journal}{Proc. Natl. Acad. Sci. USA} \bibinfo{volume}{17}
  (\bibinfo{year}{1931}) \bibinfo{pages}{315--318}. \URLprefix
  \url{https://pnas.org/doi/full/10.1073/pnas.17.5.315}.
  \DOIprefix\doi{10.1073/pnas.17.5.315}.
\bibitem[{von Neumann(1932)}]{8}
\bibinfo{author}{J.~von Neumann},
\newblock \bibinfo{title}{Zur {Operatorenmethode} {In} {Der} {Klassischen}
  {Mechanik}},
\newblock \bibinfo{journal}{The Annals of Mathematics} \bibinfo{volume}{33}
  (\bibinfo{year}{1932}) \bibinfo{pages}{587--642}. \URLprefix
  \url{https://www.jstor.org/stable/1968537?origin=crossref}.
  \DOIprefix\doi{10.2307/1968537}.
\bibitem[{Mauro(2003)}]{Mauro}
\bibinfo{author}{D.~Mauro}, \bibinfo{title}{Topics in koopman-von neumann
  theory}, \bibinfo{year}{2003}. \URLprefix
  \url{https://arxiv.org/abs/quant-ph/0301172}.
  \DOIprefix\doi{10.48550/ARXIV.QUANT-PH/0301172}.
\bibitem[{Mauro(2002)}]{Mauro:2001rm}
\bibinfo{author}{D.~Mauro},
\newblock \bibinfo{title}{{On Koopman-von Neumann waves}},
\newblock \bibinfo{journal}{Int. J. Mod. Phys. A} \bibinfo{volume}{17}
  (\bibinfo{year}{2002}) \bibinfo{pages}{1301--1325}. \URLprefix
  \url{https://doi.org/10.1142/S0217751X02009680}.
  \DOIprefix\doi{10.1142/S0217751X02009680}.
  \href{http://arxiv.org/abs/quant-ph/0105112}{{\tt arXiv:quant-ph/0105112}}.
\bibitem[{Gozzi and Mauro(2004)}]{Gozzi:2003sh}
\bibinfo{author}{E.~Gozzi}, \bibinfo{author}{D.~Mauro},
\newblock \bibinfo{title}{{On Koopman-von Neumann waves 2}},
\newblock \bibinfo{journal}{Int. J. Mod. Phys. A} \bibinfo{volume}{19}
  (\bibinfo{year}{2004}) \bibinfo{pages}{1475--1494}. \URLprefix
  \url{https://doi.org/10.1142/S0217751X04017872}.
  \DOIprefix\doi{10.1142/S0217751X04017872}.
  \href{http://arxiv.org/abs/quant-ph/0306029}{{\tt arXiv:quant-ph/0306029}}.
\bibitem[{Gozzi and Mauro(2002)}]{Gozzi:2001he}
\bibinfo{author}{E.~Gozzi}, \bibinfo{author}{D.~Mauro},
\newblock \bibinfo{title}{{Minimal coupling in Koopman-von Neumann theory}},
\newblock \bibinfo{journal}{Annals Phys.} \bibinfo{volume}{296}
  (\bibinfo{year}{2002}) \bibinfo{pages}{152--186}. \URLprefix
  \url{https://doi.org/10.1006/aphy.2001.6206}.
  \DOIprefix\doi{10.1006/aphy.2001.6206}.
  \href{http://arxiv.org/abs/quant-ph/0105113}{{\tt arXiv:quant-ph/0105113}}.
\bibitem[{Carta et~al.(2006)Carta, Gozzi, and Mauro}]{Carta:2005fq}
\bibinfo{author}{P.~Carta}, \bibinfo{author}{E.~Gozzi},
  \bibinfo{author}{D.~Mauro},
\newblock \bibinfo{title}{{Koopman-von Neumann formulation of classical
  Yang-Mills theories. I.}},
\newblock \bibinfo{journal}{Annalen Phys.} \bibinfo{volume}{15}
  (\bibinfo{year}{2006}) \bibinfo{pages}{177--215}. \URLprefix
  \url{https://doi.org/10.1002/andp.200510177}.
  \DOIprefix\doi{10.1002/andp.200510177}.
  \href{http://arxiv.org/abs/hep-th/0508244}{{\tt arXiv:hep-th/0508244}}.
\bibitem[{Abrikosov et~al.(2005)Abrikosov, Gozzi, and Mauro}]{Abrikosov:2004cf}
\bibinfo{author}{A.~A. Abrikosov, Jr.}, \bibinfo{author}{E.~Gozzi},
  \bibinfo{author}{D.~Mauro},
\newblock \bibinfo{title}{{Geometric dequantization}},
\newblock \bibinfo{journal}{Annals Phys.} \bibinfo{volume}{317}
  (\bibinfo{year}{2005}) \bibinfo{pages}{24--71}. \URLprefix
  \url{https://doi.org/10.1016/j.aop.2004.12.001}.
  \DOIprefix\doi{10.1016/j.aop.2004.12.001}.
  \href{http://arxiv.org/abs/quant-ph/0406028}{{\tt arXiv:quant-ph/0406028}}.
\bibitem[{Giulini et~al.(1996)Giulini, Kiefer, Joos, Kupsch, Stamatescu, and
  Zeh}]{Giulini:1996nw}
\bibinfo{author}{D.~Giulini}, \bibinfo{author}{C.~Kiefer},
  \bibinfo{author}{E.~Joos}, \bibinfo{author}{J.~Kupsch},
  \bibinfo{author}{I.~O. Stamatescu}, \bibinfo{author}{H.~D. Zeh},
  \bibinfo{title}{{Decoherence and the appearance of a classical world in
  quantum theory}}, \bibinfo{publisher}{Springer}, \bibinfo{address}{Berlin},
  \bibinfo{year}{1996}. \URLprefix
  \url{https://doi.org/10.1007/978-3-662-05328-7}.
\bibitem[{Padmanabhan(2015)}]{Padmanabhan_2015}
\bibinfo{author}{T.~Padmanabhan}, \bibinfo{title}{Sleeping Beauties in
  Theoretical Physics}, \bibinfo{publisher}{Springer International Publishing},
  \bibinfo{address}{Heidelberg}, \bibinfo{year}{2015}. \URLprefix
  \url{https://doi.org/10.1007\%2F978-3-319-13443-7}.
  \DOIprefix\doi{10.1007/978-3-319-13443-7}.
\bibitem[{Shankar(1994)}]{Shankar_1994}
\bibinfo{author}{R.~Shankar}, \bibinfo{title}{Principles of Quantum Mechanics},
  \bibinfo{publisher}{Springer {US}}, \bibinfo{address}{New York},
  \bibinfo{year}{1994}. \URLprefix
  \url{https://doi.org/10.1007\%2F978-1-4757-0576-8}.
  \DOIprefix\doi{10.1007/978-1-4757-0576-8}.
\bibitem[{Schlosshauer(2019)}]{Schlosshauer:2019ewh}
\bibinfo{author}{M.~Schlosshauer},
\newblock \bibinfo{title}{{Quantum Decoherence}},
\newblock \bibinfo{journal}{Phys. Rept.} \bibinfo{volume}{831}
  (\bibinfo{year}{2019}) \bibinfo{pages}{1--57}. \URLprefix
  \url{https://doi.org/10.1016/j.physrep.2019.10.001}.
  \DOIprefix\doi{10.1016/j.physrep.2019.10.001}.
  \href{http://arxiv.org/abs/1911.06282}{{\tt arXiv:1911.06282}}.
\bibitem[{Ball(2008)}]{Ball2008}
\bibinfo{author}{P.~Ball},
\newblock \bibinfo{title}{Physics: Quantum all the way},
\newblock \bibinfo{journal}{Nature} \bibinfo{volume}{453}
  (\bibinfo{year}{2008}) \bibinfo{pages}{22--25}. \URLprefix
  \url{https://doi.org/10.1038/453022a}. \DOIprefix\doi{10.1038/453022a}.
\bibitem[{Klein(2012)}]{Klein2012}
\bibinfo{author}{U.~Klein},
\newblock \bibinfo{title}{What is the limit
  $\hslash${\hspace{0.167em}}$\rightarrow${\hspace{0.167em}}0 of quantum
  theory?},
\newblock \bibinfo{journal}{Am. J. Phys.} \bibinfo{volume}{80}
  (\bibinfo{year}{2012}) \bibinfo{pages}{1009--1016}. \URLprefix
  \url{https://doi.org/10.1119/1.4751274}. \DOIprefix\doi{10.1119/1.4751274}.
\bibitem[{Holland(1993)}]{holland_1993}
\bibinfo{author}{P.~R. Holland}, \bibinfo{title}{The Quantum Theory of Motion:
  An Account of the de Broglie-Bohm Causal Interpretation of Quantum
  Mechanics}, \bibinfo{publisher}{Cambridge University Press},
  \bibinfo{address}{Cambridge}, \bibinfo{year}{1993}. \URLprefix
  \url{https://doi.org/10.1017/CBO9780511622687}.
  \DOIprefix\doi{10.1017/CBO9780511622687}.
\bibitem[{Bohm and Hiley(1993)}]{Bohm_1993}
\bibinfo{author}{D.~Bohm}, \bibinfo{author}{B.~Hiley}, \bibinfo{title}{The
  undivided universe : an ontological interpretation of quantum theory},
  \bibinfo{publisher}{Routledge}, \bibinfo{address}{London},
  \bibinfo{year}{1993}. \URLprefix \url{https://doi.org/10.4324/9780203980385}.
  \DOIprefix\doi{10.4324/9780203980385}.
\bibitem[{Berry and Mount(1972)}]{Berry:1972na}
\bibinfo{author}{M.~V. Berry}, \bibinfo{author}{K.~E. Mount},
\newblock \bibinfo{title}{{Semiclassical approximations in wave mechanics}},
\newblock \bibinfo{journal}{Rept. Prog. Phys.} \bibinfo{volume}{35}
  (\bibinfo{year}{1972}) \bibinfo{pages}{315}. \URLprefix
  \url{https://doi.org/10.1088/0034-4885/35/1/306}.
  \DOIprefix\doi{10.1088/0034-4885/35/1/306}.
\bibitem[{Vedenyapin and Fimin(2012)}]{Vedenyapin}
\bibinfo{author}{V.~Vedenyapin}, \bibinfo{author}{N.~Fimin},
\newblock \bibinfo{title}{The liouville equation, the hydrodynamic
  substitution, and the hamilton-jacobi equation},
\newblock \bibinfo{journal}{Dokl. Math.} \bibinfo{volume}{86}
  (\bibinfo{year}{2012}) \bibinfo{pages}{697--699}. \URLprefix
  \url{https://doi.org/10.1134/S1064562412050134}.
  \DOIprefix\doi{10.1134/S1064562412050134}.
\bibitem[{Sen et~al.(2020)Sen, Dhasmana, and Silagadze}]{Sen_2020}
\bibinfo{author}{A.~Sen}, \bibinfo{author}{S.~Dhasmana}, \bibinfo{author}{Z.~K.
  Silagadze},
\newblock \bibinfo{title}{Free fall in {KvN} mechanics and einstein's principle
  of equivalence},
\newblock \bibinfo{journal}{Annals of Physics} \bibinfo{volume}{422}
  (\bibinfo{year}{2020}) \bibinfo{pages}{168302}. \URLprefix
  \url{https://doi.org/10.1016\%2Fj.aop.2020.168302}.
  \DOIprefix\doi{10.1016/j.aop.2020.168302}.
\bibitem[{Ballentine(1972)}]{Ballentine}
\bibinfo{author}{L.~E. Ballentine},
\newblock \bibinfo{title}{Einstein's interpretation of quantum mechanics},
\newblock \bibinfo{journal}{Am. J. Phys.} \bibinfo{volume}{40}
  (\bibinfo{year}{1972}) \bibinfo{pages}{1763--1771}. \URLprefix
  \url{https://doi.org/10.1119/1.1987060}. \DOIprefix\doi{10.1119/1.1987060}.
\bibitem[{Einstein(1936)}]{Einstein_1936}
\bibinfo{author}{A.~Einstein},
\newblock \bibinfo{title}{Physics and reality},
\newblock \bibinfo{journal}{J. Franklin Institute} \bibinfo{volume}{221}
  (\bibinfo{year}{1936}) \bibinfo{pages}{349--382}. \URLprefix
  \url{https://doi.org/10.1016/S0016-0032(36)91047-5}.
  \DOIprefix\doi{https://doi.org/10.1016/S0016-0032(36)91047-5}.
\bibitem[{Zurek(2007)}]{Zurek2007}
\bibinfo{author}{W.~H. Zurek},
\newblock \bibinfo{title}{Decoherence and the transition from quantum to
  classical --- revisited},
\newblock in: \bibinfo{editor}{B.~Duplantier}, \bibinfo{editor}{J.-M. Raimond},
  \bibinfo{editor}{V.~Rivasseau} (Eds.), \bibinfo{booktitle}{Quantum
  Decoherence: Poincar{\'e} Seminar 2005}, \bibinfo{publisher}{Birkh{\"a}user
  Basel}, \bibinfo{address}{Basel}, \bibinfo{year}{2007}, pp.
  \bibinfo{pages}{1--31}. \URLprefix
  \url{https://doi.org/10.1007/978-3-7643-7808-0_1}.
  \DOIprefix\doi{10.1007/978-3-7643-7808-0_1}.
\bibitem[{Klein(2018)}]{Klein_2018}
\bibinfo{author}{U.~Klein},
\newblock \bibinfo{title}{From koopman–von neumann theory to quantum theory},
\newblock \bibinfo{journal}{Quantum Stud.: Math. Found.} \bibinfo{volume}{5}
  (\bibinfo{year}{2018}) \bibinfo{pages}{219--227}. \URLprefix
  \url{https://doi.org/10.1007/s40509-017-0113-2}.
  \DOIprefix\doi{10.1007/s40509-017-0113-2}.
\bibitem[{Bondar et~al.(2013)Bondar, Cabrera, Zhdanov, and
  Rabitz}]{Bondar_2013}
\bibinfo{author}{D.~I. Bondar}, \bibinfo{author}{R.~Cabrera},
  \bibinfo{author}{D.~V. Zhdanov}, \bibinfo{author}{H.~A. Rabitz},
\newblock \bibinfo{title}{Wigner phase-space distribution as a wave function},
\newblock \bibinfo{journal}{Phys. Rev. A} \bibinfo{volume}{88}
  (\bibinfo{year}{2013}) \bibinfo{pages}{052108}. \URLprefix
  \url{https://doi.org/10.1103\%2Fphysreva.88.052108}.
  \DOIprefix\doi{10.1103/physreva.88.052108}.
\bibitem[{Bondar et~al.(2012)Bondar, Cabrera, Lompay, Ivanov, and
  Rabitz}]{Bondar_2012a}
\bibinfo{author}{D.~I. Bondar}, \bibinfo{author}{R.~Cabrera},
  \bibinfo{author}{R.~R. Lompay}, \bibinfo{author}{M.~Y. Ivanov},
  \bibinfo{author}{H.~A. Rabitz},
\newblock \bibinfo{title}{Operational dynamic modeling transcending quantum and
  classical mechanics},
\newblock \bibinfo{journal}{Phys. Rev. Lett.} \bibinfo{volume}{109}
  (\bibinfo{year}{2012}) \bibinfo{pages}{190403}. \URLprefix
  \url{https://link.aps.org/doi/10.1103/PhysRevLett.109.190403}.
  \DOIprefix\doi{10.1103/PhysRevLett.109.190403}.
\bibitem[{Alvarez et~al.(2008)Alvarez, Gomis, Kamimura, and
  Plyushchay}]{Alvarez:2007ys}
\bibinfo{author}{P.~D. Alvarez}, \bibinfo{author}{J.~Gomis},
  \bibinfo{author}{K.~Kamimura}, \bibinfo{author}{M.~S. Plyushchay},
\newblock \bibinfo{title}{{Anisotropic harmonic oscillator, non-commutative
  Landau problem and exotic Newton-Hooke symmetry}},
\newblock \bibinfo{journal}{Phys. Lett. B} \bibinfo{volume}{659}
  (\bibinfo{year}{2008}) \bibinfo{pages}{906--912}. \URLprefix
  \url{https://doi.org/10.1016/j.physletb.2007.12.016}.
  \DOIprefix\doi{10.1016/j.physletb.2007.12.016}.
  \href{http://arxiv.org/abs/0711.2644}{{\tt arXiv:0711.2644}}.
\bibitem[{Alvarez et~al.(2007)Alvarez, Gomis, Kamimura, and
  Plyushchay}]{Alvarez:2007fw}
\bibinfo{author}{P.~D. Alvarez}, \bibinfo{author}{J.~Gomis},
  \bibinfo{author}{K.~Kamimura}, \bibinfo{author}{M.~S. Plyushchay},
\newblock \bibinfo{title}{{(2+1)D Exotic Newton-Hooke Symmetry, Duality and
  Projective Phase}},
\newblock \bibinfo{journal}{Annals Phys.} \bibinfo{volume}{322}
  (\bibinfo{year}{2007}) \bibinfo{pages}{1556--1586}. \URLprefix
  \url{https://doi.org/10.1016/j.aop.2007.03.002}.
  \DOIprefix\doi{10.1016/j.aop.2007.03.002}.
  \href{http://arxiv.org/abs/hep-th/0702014}{{\tt arXiv:hep-th/0702014}}.
\bibitem[{Zhang and Horvathy(2012)}]{Zhang:2011zua}
\bibinfo{author}{P.-M. Zhang}, \bibinfo{author}{P.~A. Horvathy},
\newblock \bibinfo{title}{{Chiral Decomposition in the Non-Commutative Landau
  Problem}},
\newblock \bibinfo{journal}{Annals Phys.} \bibinfo{volume}{327}
  (\bibinfo{year}{2012}) \bibinfo{pages}{1730--1743}. \URLprefix
  \url{https://doi.org/10.1016/j.aop.2012.02.014}.
  \DOIprefix\doi{10.1016/j.aop.2012.02.014}.
  \href{http://arxiv.org/abs/1112.0409}{{\tt arXiv:1112.0409}}.
\bibitem[{de~Gosson(2011)}]{deGosson2011}
\bibinfo{author}{M.~A. de~Gosson}, \bibinfo{title}{Bopp Pseudo-differential
  Operators}, \bibinfo{publisher}{Springer Basel}, \bibinfo{address}{Basel},
  \bibinfo{year}{2011}, pp. \bibinfo{pages}{291--306}. \URLprefix
  \url{https://doi.org/10.1007/978-3-7643-9992-4_18}.
  \DOIprefix\doi{10.1007/978-3-7643-9992-4_18}.
\bibitem[{Hillery et~al.(1984)Hillery, O'Connell, Scully, and
  Wigner}]{Hillery1984}
\bibinfo{author}{M.~Hillery}, \bibinfo{author}{R.~O'Connell},
  \bibinfo{author}{M.~Scully}, \bibinfo{author}{E.~Wigner},
\newblock \bibinfo{title}{Distribution functions in physics: Fundamentals},
\newblock \bibinfo{journal}{Phys. Rep.} \bibinfo{volume}{106}
  (\bibinfo{year}{1984}) \bibinfo{pages}{121--167}. \URLprefix
  \url{https://www.sciencedirect.com/science/article/pii/0370157384901601}.
  \DOIprefix\doi{10.1016/0370-1573(84)90160-1}.
\bibitem[{Sudarshan(1976)}]{Sudarshan1976}
\bibinfo{author}{E.~C.~G. Sudarshan},
\newblock \bibinfo{title}{Interaction between classical and quantum systems and
  the measurement of quantum observables},
\newblock \bibinfo{journal}{Pramana} \bibinfo{volume}{6} (\bibinfo{year}{1976})
  \bibinfo{pages}{117--126}. \URLprefix
  \url{https://doi.org/10.1007/bf02847120}. \DOIprefix\doi{10.1007/bf02847120}.
\bibitem[{Sherry and Sudarshan(1978)}]{Sherry1978}
\bibinfo{author}{T.~N. Sherry}, \bibinfo{author}{E.~C.~G. Sudarshan},
\newblock \bibinfo{title}{Interaction between classical and quantum systems: A
  new approach to quantum measurement.i},
\newblock \bibinfo{journal}{Phys. Rev. D} \bibinfo{volume}{18}
  (\bibinfo{year}{1978}) \bibinfo{pages}{4580--4589}. \URLprefix
  \url{https://doi.org/10.1103/physrevd.18.4580}.
  \DOIprefix\doi{10.1103/physrevd.18.4580}.
\bibitem[{Morgan(2020)}]{Morgan:2019azd}
\bibinfo{author}{P.~Morgan},
\newblock \bibinfo{title}{{An algebraic approach to Koopman classical
  mechanics}},
\newblock \bibinfo{journal}{Annals Phys.} \bibinfo{volume}{414}
  (\bibinfo{year}{2020}) \bibinfo{pages}{168090}. \URLprefix
  \url{https://doi.org/10.1016/j.aop.2020.168090}.
  \DOIprefix\doi{10.1016/j.aop.2020.168090}.
  \href{http://arxiv.org/abs/1901.00526}{{\tt arXiv:1901.00526}}.
\bibitem[{Chashchina et~al.(2020)Chashchina, Sen, and
  Silagadze}]{Chashchina_2020}
\bibinfo{author}{O.~I. Chashchina}, \bibinfo{author}{A.~Sen},
  \bibinfo{author}{Z.~K. Silagadze},
\newblock \bibinfo{title}{On deformations of classical mechanics due to
  planck-scale physics},
\newblock \bibinfo{journal}{Int. J. Mod. Phys. D} \bibinfo{volume}{29}
  (\bibinfo{year}{2020}) \bibinfo{pages}{2050070}. \URLprefix
  \url{https://doi.org/10.1142/s0218271820500704}.
  \DOIprefix\doi{10.1142/s0218271820500704}.
\bibitem[{Wilczek(2022)}]{wilczek_2022}
\bibinfo{author}{F.~Wilczek}, \bibinfo{title}{Notes on koopman von neumann
  mechanics, and a step beyond}, \bibinfo{year}{2022}. \URLprefix
  \url{http://frankwilczek.com/2015/koopmanVonNeumann02.pdf}.
\bibitem[{Havas(1964)}]{Havas}
\bibinfo{author}{P.~Havas},
\newblock \bibinfo{title}{Four-dimensional formulations of newtonian mechanics
  and their relation to the special and the general theory of relativity},
\newblock \bibinfo{journal}{Rev. Mod. Phys.} \bibinfo{volume}{36}
  (\bibinfo{year}{1964}) \bibinfo{pages}{938--965}. \URLprefix
  \url{https://doi.org/10.1103/RevModPhys.36.938}.
  \DOIprefix\doi{10.1103/RevModPhys.36.938}.
\bibitem[{Cartan(1923)}]{Cartan1923}
\bibinfo{author}{E.~Cartan},
\newblock \bibinfo{title}{Sur les vari{\'{e}}t{\'{e}}s {\`{a}} connexion affine
  et la th{\'{e}}orie de la relativit{\'{e}} g{\'{e}}n{\'{e}}ralis{\'{e}}e
  (premi{\`{e}}re partie)},
\newblock \bibinfo{journal}{Annales scientifiques de l\'{E}cole normale
  sup\'{e}rieure} \bibinfo{volume}{40} (\bibinfo{year}{1923})
  \bibinfo{pages}{325--412}. \URLprefix
  \url{https://doi.org/10.24033/asens.751}. \DOIprefix\doi{10.24033/asens.751}.
\bibitem[{Andringa et~al.(2011)Andringa, Bergshoeff, Panda, and
  de~Roo}]{Andringa:2010it}
\bibinfo{author}{R.~Andringa}, \bibinfo{author}{E.~Bergshoeff},
  \bibinfo{author}{S.~Panda}, \bibinfo{author}{M.~de~Roo},
\newblock \bibinfo{title}{{Newtonian Gravity and the Bargmann Algebra}},
\newblock \bibinfo{journal}{Class. Quant. Grav.} \bibinfo{volume}{28}
  (\bibinfo{year}{2011}) \bibinfo{pages}{105011}. \URLprefix
  \url{https://doi.org/10.1088/0264-9381/28/10/105011}.
  \DOIprefix\doi{10.1088/0264-9381/28/10/105011}.
  \href{http://arxiv.org/abs/1011.1145}{{\tt arXiv:1011.1145}}.
\bibitem[{Kirillov(1962)}]{Kirillov_1962}
\bibinfo{author}{A.~Kirillov},
\newblock \bibinfo{title}{{Unitary} {representations} {of} {nilpotent} {Lie}
  {groups}},
\newblock \bibinfo{journal}{Russ. Math. Surv.} \bibinfo{volume}{17}
  (\bibinfo{year}{1962}) \bibinfo{pages}{53--104}. \URLprefix
  \url{https://doi.org/10.1070/rm1962v017n04abeh004118}.
  \DOIprefix\doi{10.1070/rm1962v017n04abeh004118}.
\bibitem[{Kirillov(2001)}]{Kirillov2001}
\bibinfo{author}{A.~A. Kirillov},
\newblock \bibinfo{title}{Geometric quantization},
\newblock in: \bibinfo{editor}{V.~I. Arnold}, \bibinfo{editor}{S.~P. Novikov}
  (Eds.), \bibinfo{booktitle}{Dynamical Systems IV: Symplectic Geometry and its
  Applications}, \bibinfo{publisher}{Springer Berlin Heidelberg},
  \bibinfo{address}{Berlin, Heidelberg}, \bibinfo{year}{2001}, pp.
  \bibinfo{pages}{139--176}. \URLprefix
  \url{https://doi.org/10.1007/978-3-662-06791-8_2}.
  \DOIprefix\doi{10.1007/978-3-662-06791-8_2}.
\bibitem[{Kostant(1970)}]{Kostant}
\bibinfo{author}{B.~Kostant},
\newblock \bibinfo{title}{Quantization and unitary representations},
\newblock in: \bibinfo{editor}{C.~Taam} (Ed.), \bibinfo{booktitle}{Lectures in
  Modern Analysis and Applications III}, \bibinfo{publisher}{Springer},
  \bibinfo{address}{Berlin}, \bibinfo{year}{1970}, pp.
  \bibinfo{pages}{87--208}. \URLprefix
  \url{https://doi.org/10.1007/BFb0079068}. \DOIprefix\doi{10.1007/BFb0079068}.
\bibitem[{Souriau(1970)}]{Souriau}
\bibinfo{author}{J.-M. Souriau}, \bibinfo{title}{Structure des syst\'{e}mes
  dynamiques}, \bibinfo{publisher}{Dunod}, \bibinfo{address}{Paris},
  \bibinfo{year}{1970}. \URLprefix
  \url{https://link.springer.com/book/9780817636951}.
\bibitem[{Souriau(1966)}]{Souriau1}
\bibinfo{author}{J.-M. Souriau},
\newblock \bibinfo{title}{{Quantification g\'{e}om\'{e}trique}},
\newblock \bibinfo{journal}{Comm. Math. Phys.} \bibinfo{volume}{1}
  (\bibinfo{year}{1966}) \bibinfo{pages}{374--398}. \URLprefix
  \url{https://doi.org/cmp/1103758996}. \DOIprefix\doi{cmp/1103758996}.
\bibitem[{\'{S}niatycki(2016)}]{Sniatycki}
\bibinfo{author}{J.~\'{S}niatycki},
\newblock \bibinfo{title}{Lectures on geometric quantization},
\newblock in: \bibinfo{editor}{I.~M. Mladenov}, \bibinfo{editor}{G.~Mengand},
  \bibinfo{editor}{A.~Yoshioka} (Eds.), \bibinfo{booktitle}{Seventeenth
  International Conference onGeometry, Integrability and Quantization},
  \bibinfo{publisher}{Avangard Prima}, \bibinfo{address}{Sofia},
  \bibinfo{year}{2016}, pp. \bibinfo{pages}{95--129}. \URLprefix
  \url{https://doi.org/10.7546/giq-17-2016-95-129}.
  \DOIprefix\doi{10.7546/giq-17-2016-95-129}.
\bibitem[{Abrikosov et~al.(2003)Abrikosov, Gozzi, and Mauro}]{Abrikosov:2003ce}
\bibinfo{author}{A.~A. Abrikosov, Jr.}, \bibinfo{author}{E.~Gozzi},
  \bibinfo{author}{D.~Mauro},
\newblock \bibinfo{title}{{Time and geometric quantization}},
\newblock \bibinfo{journal}{Mod. Phys. Lett. A} \bibinfo{volume}{18}
  (\bibinfo{year}{2003}) \bibinfo{pages}{2347--2354}. \URLprefix
  \url{https://doi.org/10.1142/S0217732303012568}.
  \DOIprefix\doi{10.1142/S0217732303012568}.
  \href{http://arxiv.org/abs/quant-ph/0308101}{{\tt arXiv:quant-ph/0308101}}.
\bibitem[{Bondar et~al.(2019)Bondar, Gay-Balmaz, and Tronci}]{Bondar_2019}
\bibinfo{author}{D.~I. Bondar}, \bibinfo{author}{F.~Gay-Balmaz},
  \bibinfo{author}{C.~Tronci},
\newblock \bibinfo{title}{Koopman wavefunctions and classical-quantum
  correlation dynamics},
\newblock \bibinfo{journal}{Proc. R. Soc. A} \bibinfo{volume}{475}
  (\bibinfo{year}{2019}) \bibinfo{pages}{20180879}. \URLprefix
  \url{https://doi.org/10.1098/rspa.2018.0879}.
  \DOIprefix\doi{10.1098/rspa.2018.0879}.
\bibitem[{Kunzle and Duval(1986)}]{Kunzle1986}
\bibinfo{author}{H.~P. Kunzle}, \bibinfo{author}{C.~Duval},
\newblock \bibinfo{title}{Relativistic and non-relativistic classical field
  theory on five-dimensional spacetime},
\newblock \bibinfo{journal}{Class. Quant. Grav.} \bibinfo{volume}{3}
  (\bibinfo{year}{1986}) \bibinfo{pages}{957--974}. \URLprefix
  \url{https://doi.org/10.1088/0264-9381/3/5/024}.
  \DOIprefix\doi{10.1088/0264-9381/3/5/024}.
\bibitem[{Gibbons et~al.(2011)Gibbons, Houri, Kubiznak, and
  Warnick}]{Gibbons:2011hg}
\bibinfo{author}{G.~W. Gibbons}, \bibinfo{author}{T.~Houri},
  \bibinfo{author}{D.~Kubiznak}, \bibinfo{author}{C.~M. Warnick},
\newblock \bibinfo{title}{{Some Spacetimes with Higher Rank Killing-Stackel
  Tensors}},
\newblock \bibinfo{journal}{Phys. Lett. B} \bibinfo{volume}{700}
  (\bibinfo{year}{2011}) \bibinfo{pages}{68--74}. \URLprefix
  \url{https://link.aps.org/doi/10.1016/j.physletb.2011.04.047}.
  \DOIprefix\doi{10.1016/j.physletb.2011.04.047}.
  \href{http://arxiv.org/abs/1103.5366}{{\tt arXiv:1103.5366}}.
\bibitem[{Cariglia and Alves(2015)}]{Cariglia_2015}
\bibinfo{author}{M.~Cariglia}, \bibinfo{author}{F.~K. Alves},
\newblock \bibinfo{title}{The eisenhart lift: a didactical introduction of
  modern geometrical concepts from hamiltonian dynamics},
\newblock \bibinfo{journal}{Eur. J. Phys.} \bibinfo{volume}{36}
  (\bibinfo{year}{2015}) \bibinfo{pages}{025018}. \URLprefix
  \url{https://doi.org/10.1088/0143-0807/36/2/025018}.
  \DOIprefix\doi{10.1088/0143-0807/36/2/025018}.
\bibitem[{Arnowitt et~al.(2008)Arnowitt, Deser, and Misner}]{Arnowitt:1962hi}
\bibinfo{author}{R.~L. Arnowitt}, \bibinfo{author}{S.~Deser},
  \bibinfo{author}{C.~W. Misner},
\newblock \bibinfo{title}{{The Dynamics of general relativity}},
\newblock \bibinfo{journal}{Gen. Rel. Grav.} \bibinfo{volume}{40}
  (\bibinfo{year}{2008}) \bibinfo{pages}{1997--2027}. \URLprefix
  \url{https://doi.org/10.1007/s10714-008-0661-1}.
  \DOIprefix\doi{10.1007/s10714-008-0661-1}.
  \href{http://arxiv.org/abs/gr-qc/0405109}{{\tt arXiv:gr-qc/0405109}}.
\bibitem[{Duval et~al.(2014)Duval, Gibbons, Horvathy, and
  Zhang}]{Duval:2014uoa}
\bibinfo{author}{C.~Duval}, \bibinfo{author}{G.~W. Gibbons},
  \bibinfo{author}{P.~A. Horvathy}, \bibinfo{author}{P.~M. Zhang},
\newblock \bibinfo{title}{{Carroll versus Newton and Galilei: two dual
  non-Einsteinian concepts of time}},
\newblock \bibinfo{journal}{Class. Quant. Grav.} \bibinfo{volume}{31}
  (\bibinfo{year}{2014}) \bibinfo{pages}{085016}. \URLprefix
  \url{https://doi.org/10.1088/0264-9381/31/8/085016}.
  \DOIprefix\doi{10.1088/0264-9381/31/8/085016}.
  \href{http://arxiv.org/abs/1402.0657}{{\tt arXiv:1402.0657}}.
\bibitem[{Zhang et~al.(2020)Zhang, Cariglia, Elbistan, and
  Horvathy}]{Zhang:2019gdm}
\bibinfo{author}{P.~M. Zhang}, \bibinfo{author}{M.~Cariglia},
  \bibinfo{author}{M.~Elbistan}, \bibinfo{author}{P.~A. Horvathy},
\newblock \bibinfo{title}{{Scaling and conformal symmetries for plane
  gravitational waves}},
\newblock \bibinfo{journal}{J. Math. Phys.} \bibinfo{volume}{61}
  (\bibinfo{year}{2020}) \bibinfo{pages}{022502}. \URLprefix
  \url{https://doi.org/10.1063/1.5136078}. \DOIprefix\doi{10.1063/1.5136078}.
  \href{http://arxiv.org/abs/1905.08661}{{\tt arXiv:1905.08661}}.
\bibitem[{N\v{a}stase(2019)}]{Nastase}
\bibinfo{author}{H.~N\v{a}stase}, \bibinfo{title}{Introduction to Quantum Field
  Theory}, \bibinfo{publisher}{Cambridge University Press},
  \bibinfo{address}{Cambridge}, \bibinfo{year}{2019}. \URLprefix
  \url{https://books.google.ru/books?id=mm-rDwAAQBAJ}.
\bibitem[{Horvathy and Zhang(2009)}]{Horvathy:2008hd}
\bibinfo{author}{P.~A. Horvathy}, \bibinfo{author}{P.~Zhang},
\newblock \bibinfo{title}{{Vortices in (abelian) Chern-Simons gauge theory}},
\newblock \bibinfo{journal}{Phys. Rept.} \bibinfo{volume}{481}
  (\bibinfo{year}{2009}) \bibinfo{pages}{83--142}. \URLprefix
  \url{https://doi.org/10.1016/j.physrep.2009.07.003}.
  \DOIprefix\doi{10.1016/j.physrep.2009.07.003}.
  \href{http://arxiv.org/abs/0811.2094}{{\tt arXiv:0811.2094}}.
\bibitem[{Minguzzi(2006)}]{Minguzzi:2006wz}
\bibinfo{author}{E.~Minguzzi},
\newblock \bibinfo{title}{{Classical aspects of lightlike dimensional
  reduction}},
\newblock \bibinfo{journal}{Class. Quant. Grav.} \bibinfo{volume}{23}
  (\bibinfo{year}{2006}) \bibinfo{pages}{7085--7110}. \URLprefix
  \url{https://doi.org/10.1088/0264-9381/23/23/029}.
  \DOIprefix\doi{10.1088/0264-9381/23/23/029}.
  \href{http://arxiv.org/abs/gr-qc/0610011}{{\tt arXiv:gr-qc/0610011}}.
\bibitem[{Duval et~al.(1994)Duval, Horvathy, and Palla}]{Duval:1994qye}
\bibinfo{author}{C.~Duval}, \bibinfo{author}{P.~A. Horvathy},
  \bibinfo{author}{L.~Palla},
\newblock \bibinfo{title}{{Conformal Properties of Chern-Simons Vortices in
  External Fields}},
\newblock \bibinfo{journal}{Phys. Rev. D} \bibinfo{volume}{50}
  (\bibinfo{year}{1994}) \bibinfo{pages}{6658--6661}. \URLprefix
  \url{https://doi.org/10.1103/PhysRevD.50.6658}.
  \DOIprefix\doi{10.1103/PhysRevD.50.6658}.
  \href{http://arxiv.org/abs/hep-th/9404047}{{\tt arXiv:hep-th/9404047}}.
\bibitem[{Gibbons(2014)}]{Gibbons:2014zla}
\bibinfo{author}{G.~W. Gibbons}, \bibinfo{title}{Dark energy and the schwarzian
  derivative}, \bibinfo{year}{2014}. \URLprefix
  \url{https://arxiv.org/abs/1403.5431}.
  \DOIprefix\doi{10.48550/ARXIV.1403.5431}.
  \href{http://arxiv.org/abs/1403.5431}{{\tt arXiv:1403.5431}}.
\bibitem[{Zhang et~al.(2022)Zhang, Zhao, and Horvathy}]{Zhang:2021ssp}
\bibinfo{author}{P.~Zhang}, \bibinfo{author}{Q.~Zhao}, \bibinfo{author}{P.~A.
  Horvathy},
\newblock \bibinfo{title}{{Gravitational waves and conformal time
  transformations}},
\newblock \bibinfo{journal}{Annals Phys.} \bibinfo{volume}{440}
  (\bibinfo{year}{2022}) \bibinfo{pages}{168833}. \URLprefix
  \url{https://doi.org/10.1016/j.aop.2022.168833}.
  \DOIprefix\doi{10.1016/j.aop.2022.168833}.
  \href{http://arxiv.org/abs/2112.09589}{{\tt arXiv:2112.09589}}.
\bibitem[{Niederer(1973)}]{Niederer:1973tz}
\bibinfo{author}{U.~Niederer},
\newblock \bibinfo{title}{{The maximal kinematical invariance group of the
  harmonic oscillator}},
\newblock \bibinfo{journal}{Helv. Phys. Acta} \bibinfo{volume}{46}
  (\bibinfo{year}{1973}) \bibinfo{pages}{191--200}.
\bibitem[{Yariv(1967)}]{Yariv}
\bibinfo{author}{A.~Yariv}, \bibinfo{title}{Quantum Electronics},
  \bibinfo{publisher}{Wiley}, \bibinfo{address}{New York},
  \bibinfo{year}{1967}.
\bibitem[{Steuernagel(2014)}]{Steuernagel2014}
\bibinfo{author}{O.~Steuernagel},
\newblock \bibinfo{title}{Equivalence between free quantum particles and those
  in harmonic potentials and its application to instantaneous changes},
\newblock \bibinfo{journal}{Eur. Phys. J. Plus} \bibinfo{volume}{129}
  (\bibinfo{year}{2014}) \bibinfo{pages}{114}. \URLprefix
  \url{https://doi.org/10.1140/epjp/i2014-14114-3}.
  \DOIprefix\doi{10.1140/epjp/i2014-14114-3}.
\bibitem[{Solov'ev(1982)}]{Solovev:1982}
\bibinfo{author}{E.~A. Solov'ev},
\newblock \bibinfo{title}{{Connection between problems of the harmonic
  oscillator and a free particle in quantum mechanics}},
\newblock \bibinfo{journal}{Sov. J. Nucl. Phys.} \bibinfo{volume}{35}
  (\bibinfo{year}{1982}) \bibinfo{pages}{136--137}.
\bibitem[{Jackiw(1980)}]{Jackiw:1980mm}
\bibinfo{author}{R.~Jackiw},
\newblock \bibinfo{title}{{Dynamical Symmetry of the Magnetic Monopole}},
\newblock \bibinfo{journal}{Annals Phys.} \bibinfo{volume}{129}
  (\bibinfo{year}{1980}) \bibinfo{pages}{183}. \URLprefix
  \url{https://doi.org/10.1016/0003-4916(80)90295-X}.
  \DOIprefix\doi{10.1016/0003-4916(80)90295-X}.
\bibitem[{Guerrero and Lopez-Ruiz(2013)}]{Guerrero:2013mdt}
\bibinfo{author}{J.~Guerrero}, \bibinfo{author}{F.~F. Lopez-Ruiz},
\newblock \bibinfo{title}{{The Quantum Arnold Transformation and its
  applications}},
\newblock \bibinfo{journal}{Nuovo Cim. C} \bibinfo{volume}{36}
  (\bibinfo{year}{2013}) \bibinfo{pages}{127--137}.
  \DOIprefix\doi{10.1393/ncc/i2013-11528-0}.
\bibitem[{McCaul and Bondar(2022)}]{McCaul:2022cyl}
\bibinfo{author}{G.~McCaul}, \bibinfo{author}{D.~I. Bondar},
\newblock \bibinfo{title}{{Free to harmonic unitary transformations in quantum
  and Koopman dynamics}},
\newblock \bibinfo{journal}{J. Phys. A} \bibinfo{volume}{55}
  (\bibinfo{year}{2022}) \bibinfo{pages}{434003}. \URLprefix
  \url{https://doi.org/10.1088/1751-8121/ac97cf}.
  \DOIprefix\doi{10.1088/1751-8121/ac97cf}.
  \href{http://arxiv.org/abs/2207.09515}{{\tt arXiv:2207.09515}}.
\bibitem[{Drach(1935)}]{drach}
\bibinfo{author}{J.~Drach},
\newblock \bibinfo{title}{Sur l'{\guillemotleft}int{\'e}gration
  logique{\guillemotright} des {\'e}quations de la dynamique},
\newblock \bibinfo{journal}{{\v{C}}asopis pro p{\v{e}}stov{\'a}n{\'\i}
  matematiky a fysiky} \bibinfo{volume}{64} (\bibinfo{year}{1935})
  \bibinfo{pages}{141--143}. \URLprefix
  \url{https://doi.org/10.21136/CPMF.1935.121249}.
  \DOIprefix\doi{10.21136/CPMF.1935.121249}.
\bibitem[{Galajinsky(2017)}]{Galajinsky:2017qsb}
\bibinfo{author}{A.~Galajinsky},
\newblock \bibinfo{title}{{Eisenhart lift in pseudo\textendash{}Euclidean space
  and higher rank killing tensors}},
\newblock \bibinfo{journal}{Phys. Part. Nucl. Lett.} \bibinfo{volume}{14}
  (\bibinfo{year}{2017}) \bibinfo{pages}{328--330}. \URLprefix
  \url{https://doi.org/10.1134/S154747711702011X}.
  \DOIprefix\doi{10.1134/S154747711702011X}.
\bibitem[{Cariglia and Galajinsky(2015)}]{Cariglia:2015fva}
\bibinfo{author}{M.~Cariglia}, \bibinfo{author}{A.~Galajinsky},
\newblock \bibinfo{title}{{Ricci-flat spacetimes admitting higher rank Killing
  tensors}},
\newblock \bibinfo{journal}{Phys. Lett. B} \bibinfo{volume}{744}
  (\bibinfo{year}{2015}) \bibinfo{pages}{320--324}. \URLprefix
  \url{https://doi.org/10.1016/j.physletb.2015.04.001}.
  \DOIprefix\doi{10.1016/j.physletb.2015.04.001}.
  \href{http://arxiv.org/abs/1503.02162}{{\tt arXiv:1503.02162}}.
\bibitem[{Filyukov and Galajinsky(2015)}]{Filyukov:2015qna}
\bibinfo{author}{S.~Filyukov}, \bibinfo{author}{A.~Galajinsky},
\newblock \bibinfo{title}{{Self-dual metrics with maximally superintegrable
  geodesic flows}},
\newblock \bibinfo{journal}{Phys. Rev. D} \bibinfo{volume}{91}
  (\bibinfo{year}{2015}) \bibinfo{pages}{104020}. \URLprefix
  \url{https://doi.org/10.1103/PhysRevD.91.104020}.
  \DOIprefix\doi{10.1103/PhysRevD.91.104020}.
  \href{http://arxiv.org/abs/1504.03826}{{\tt arXiv:1504.03826}}.
\bibitem[{Barrett et~al.(1994)Barrett, Gibbons, Perry, Pope, and
  Ruback}]{Barrett:1993yn}
\bibinfo{author}{J.~W. Barrett}, \bibinfo{author}{G.~W. Gibbons},
  \bibinfo{author}{M.~J. Perry}, \bibinfo{author}{C.~N. Pope},
  \bibinfo{author}{P.~Ruback},
\newblock \bibinfo{title}{{Kleinian geometry and the N=2 superstring}},
\newblock \bibinfo{journal}{Int. J. Mod. Phys. A} \bibinfo{volume}{9}
  (\bibinfo{year}{1994}) \bibinfo{pages}{1457--1494}. \URLprefix
  \url{https://doi.org/10.1142/S0217751X94000650}.
  \DOIprefix\doi{10.1142/S0217751X94000650}.
  \href{http://arxiv.org/abs/hep-th/9302073}{{\tt arXiv:hep-th/9302073}}.
\bibitem[{Tegmark(1997)}]{Tegmark:1997jg}
\bibinfo{author}{M.~Tegmark},
\newblock \bibinfo{title}{{On the dimensionality of space-time}},
\newblock \bibinfo{journal}{Class. Quant. Grav.} \bibinfo{volume}{14}
  (\bibinfo{year}{1997}) \bibinfo{pages}{L69--L75}. \URLprefix
  \url{https://doi.org/10.1088/0264-9381/14/4/002}.
  \DOIprefix\doi{10.1088/0264-9381/14/4/002}.
  \href{http://arxiv.org/abs/gr-qc/9702052}{{\tt arXiv:gr-qc/9702052}}.
\bibitem[{Dehdashti et~al.(2021)Dehdashti, Shahsafi, Zheng, Shen, Wang, Zhu,
  Chen, and Chen}]{Dehdashti:2021vmz}
\bibinfo{author}{S.~Dehdashti}, \bibinfo{author}{A.~Shahsafi},
  \bibinfo{author}{B.~Zheng}, \bibinfo{author}{L.~Shen},
  \bibinfo{author}{Z.~Wang}, \bibinfo{author}{R.~Zhu},
  \bibinfo{author}{H.~Chen}, \bibinfo{author}{H.~Chen},
\newblock \bibinfo{title}{{Conformal hyperbolic optics}},
\newblock \bibinfo{journal}{Phys. Rev. Res.} \bibinfo{volume}{3}
  (\bibinfo{year}{2021}) \bibinfo{pages}{033281}. \URLprefix
  \url{https://doi.org/10.1103/PhysRevResearch.3.033281}.
  \DOIprefix\doi{10.1103/PhysRevResearch.3.033281}.
\bibitem[{Alves-J\'unior et~al.(2021)Alves-J\'unior, Barreto, and
  Moraes}]{Alves-Junior:2020nva}
\bibinfo{author}{F.~A.~P. Alves-J\'unior}, \bibinfo{author}{A.~B. Barreto},
  \bibinfo{author}{F.~Moraes},
\newblock \bibinfo{title}{{Implications of Kleinian relativity}},
\newblock \bibinfo{journal}{Phys. Rev. D} \bibinfo{volume}{103}
  (\bibinfo{year}{2021}) \bibinfo{pages}{044023}. \URLprefix
  \url{https://doi.org/10.1103/PhysRevD.103.044023}.
  \DOIprefix\doi{10.1103/PhysRevD.103.044023}.
  \href{http://arxiv.org/abs/2012.03921}{{\tt arXiv:2012.03921}}.
\bibitem[{Gibbons(2020)}]{Gibbons:2020nzu}
\bibinfo{author}{G.~W. Gibbons}, \bibinfo{title}{{Lifting the Eisenhart-Duval
  Lift to a Minimal Brane}}, \bibinfo{year}{2020}. \URLprefix
  \url{https://doi.org/10.48550/arXiv.2003.06179}.
  \href{http://arxiv.org/abs/2003.06179}{{\tt arXiv:2003.06179}}.
\bibitem[{Bars(2001)}]{Bars:2000qm}
\bibinfo{author}{I.~Bars},
\newblock \bibinfo{title}{{Survey of two time physics}},
\newblock \bibinfo{journal}{Class. Quant. Grav.} \bibinfo{volume}{18}
  (\bibinfo{year}{2001}) \bibinfo{pages}{3113--3130}. \URLprefix
  \url{https://doi.org/10.1088/0264-9381/18/16/303}.
  \DOIprefix\doi{10.1088/0264-9381/18/16/303}.
  \href{http://arxiv.org/abs/hep-th/0008164}{{\tt arXiv:hep-th/0008164}}.
\bibitem[{Sakharov(1984)}]{Sakharov:1984csx}
\bibinfo{author}{A.~D. Sakharov},
\newblock \bibinfo{title}{{Cosmological Transitions With a Change in Metric
  Signature}},
\newblock \bibinfo{journal}{Sov. Phys. JETP} \bibinfo{volume}{60}
  (\bibinfo{year}{1984}) \bibinfo{pages}{214--218}. \URLprefix
  \url{https://doi.org/10.1070/PU1991v034n05ABEH002502}.
  \DOIprefix\doi{10.1070/PU1991v034n05ABEH002502}.
\bibitem[{Smolyaninov and Narimanov(2010)}]{Smolyaninov:2010tpq}
\bibinfo{author}{I.~I. Smolyaninov}, \bibinfo{author}{E.~E. Narimanov},
\newblock \bibinfo{title}{{Metric Signature Transitions in Optical
  Metamaterials}},
\newblock \bibinfo{journal}{Phys. Rev. Lett.} \bibinfo{volume}{105}
  (\bibinfo{year}{2010}) \bibinfo{pages}{067402}. \URLprefix
  \url{https://doi.org/10.1103/PhysRevLett.105.067402}.
  \DOIprefix\doi{10.1103/PhysRevLett.105.067402}.
  \href{http://arxiv.org/abs/0908.2407}{{\tt arXiv:0908.2407}}.
\bibitem[{Figueiredo et~al.(2016)Figueiredo, Gomes, Fumeron, Berche, and
  Moraes}]{Figueiredo:2016xfc}
\bibinfo{author}{D.~Figueiredo}, \bibinfo{author}{F.~A. Gomes},
  \bibinfo{author}{S.~Fumeron}, \bibinfo{author}{B.~Berche},
  \bibinfo{author}{F.~Moraes},
\newblock \bibinfo{title}{{Modeling Kleinian cosmology with electronic
  metamaterials}},
\newblock \bibinfo{journal}{Phys. Rev. D} \bibinfo{volume}{94}
  (\bibinfo{year}{2016}) \bibinfo{pages}{044039}. \URLprefix
  \url{https://doi.org/10.1103/PhysRevD.94.044039}.
  \DOIprefix\doi{10.1103/PhysRevD.94.044039}.
  \href{http://arxiv.org/abs/1608.03812}{{\tt arXiv:1608.03812}}.

\end{thebibliography}

\end{document}